\documentclass[12pt]{article}
\usepackage{amsmath}
\usepackage{amssymb}
\usepackage{epsfig}
\usepackage[latin5]{inputenc}

\numberwithin{equation}{section}

\newcommand{\gen}[1]{\partial_{#1}}

\newcommand{\ii}{i}

\DeclareMathOperator{\Sl}{sl}

\DeclareMathOperator{\simi}{gs}

\begin{document}
\title{\bf
\large Variable coefficient nonlinear Schr\"{o}dinger equations with four-dimensional symmetry groups and analysis of their solutions}

\author{C. \"{O}zemir\footnotemark[1]\thanks{Department of Mathematics, Faculty of Science and Letters,
Istanbul Technical University, 34469 Istanbul,
Turkey, e-mail: ozemir@itu.edu.tr}, \and F. G\"{u}ng\"{o}r\thanks{Department of Mathematics, Faculty of Arts and Sciences, Do\u{g}u\c{s} University, 34722 Istanbul, Turkey, e-mail: fgungor@dogus.edu.tr}
}

\date{\today}

\maketitle

\begin{abstract}
Analytical solutions of variable coefficient nonlinear Schr\"{o}dinger equations having four-dimensional symmetry groups which are in fact the next closest to the integrable ones occurring only when the Lie symmetry group is five-dimensional are obtained using two different tools. The first tool is to use one dimensional subgroups of the full symmetry group to generate solutions from those of the reduced ODEs (Ordinary Differential Equations), namely group invariant solutions. The other is by truncation in their Painlev\'e expansions.
\end{abstract}

\section{Introduction}
 The purpose of this paper is to classify solutions of  a general class of variable coefficient nonlinear Schr\"{o}dinger equations (VCNLS) of the form
\begin{eqnarray}\label{VCS}
\begin{split}
i\psi_t+f(x,t)\psi_{xx}+g(x,t) |\psi|^2 \psi+h(x,t) \psi=0,\\
f=f_1+{i}f_2,   \quad g=g_1+{i}g_2, \quad h=h_1+{i}h_2,\\
f_j,g_j,h_j\in\mathbb{R}, \quad j=1,2, \quad f_1\neq0, \quad
g_1\neq0.
\end{split}
\end{eqnarray}
with the property that they are invariant under four-dimensional Lie symmetry groups.
This class of equations models various nonlinear phenomena, for instance see
\cite{Ozemir10} and the references therein.  Symmetry classes
of \eqref{VCS} are
obtained in \cite{Winternitz93} and canonical equations admitting Lie symmetry
algebras $L$ of dimension $1\leq \dim L \leq5$ are presented there.   A suitable
basis for the maximal algebra $L$ ($\dim L=5$) is spanned by
\begin{equation}\label{alg}
T=\partial_{t},  \quad P=\partial_{x}, \quad
W=\partial_{\omega}, \quad
B=t\partial_{x}+\frac{1}{2}x\partial_{\omega},
 \quad D=t\partial_{t}+\frac{1}{2}x\partial_{x}-\frac{1}{2}\rho\partial_{\rho},
\end{equation}
which is isomorphic to the  one-dimensional extended Galilei
similitude algebra $\simi(1)$. Here $\psi\in \mathbb{C}$ is expressed in terms
of the modulus and the phase of the wave function
\begin{equation}\label{psi}
\psi(x,t)=\rho(x,t)e^{i \omega(x,t)}.
\end{equation} An equation of class \eqref{VCS} admits this algebra as long as the coefficients
$f,g$ and $h$ can be mapped into
\begin{equation}\label{sbt}
f=1, \quad  g=\epsilon+\mathrm{i} g_2, \quad \epsilon=\pm1,\quad
g_2=\text{const.}, \quad h=0
\end{equation}
by point transformations. This is nothing but the standard
cubic  nonlinear Schr\"{o}dinger equation (NLSE). For the form of
the coefficients obeying the constraints
imposed by the Painlevé test, we had been able to transform \eqref{VCS}
to the usual NLSE. In two recent papers
\cite{Khawaja10} and \cite{Brugarino10},  the conditions imposed by the Painlevé test  were shown to be equivalent to those having a Lax pair. Therefore these conditions  are also
necessary for integrability.

We intend to present  a detailed analysis
of  solutions to VCNLS equations in the absence of integrability using two different approaches. We focus on the canonical equations which are representatives of the equations from class \eqref{VCS} having
four-dimensional Lie algebras. A list of four-dimensional symmetry
algebras and the corresponding coefficients for the
canonical equations is given in Table \ref{algebras} (We  hereby correct an error on $L_5$
in \cite{Winternitz93} and slightly improve  the form of the equation).

\begin{table}[ht]
\caption{Four dimensional symmetry algebras and the coefficients in \eqref{VCS}} % title of Table
\centering % used for centering table
\begin{tabular}{c c c c c c} % centered columns (4 columns)
\hline\hline %inserts double horizontal lines
No & Algebra  &  $f$  & $g$  &  $h$ & Conditions \\ [0.5ex] % inserts table
%heading
\hline % inserts single horizontal line
$L_1$ & $\{T,D_1,C_1,W\}$ & $1$                & $(\epsilon+i\gamma)\frac{1}{x}$              & $(h_1+ih_2)\frac{1}{x^2}$  &\\ % inserting body of the table
$L_2$ & $\{T,P,B,W\}$ & $1$                & $\epsilon+i\gamma$                           & $ih_2$                     &$h_2 \neq 0$\\
$L_3$ & $\{T,P,D_2,W\}$ & $1+if_2$  & $\epsilon+i\gamma$                           & $0$                                 &$f_2 \neq 0$ \\
$L_4$ & $\{P,B,D_2,W\}$ & $1$                & $\epsilon+i\gamma$                           & $i\frac{h_2}{t}$           &$h_2 \neq 0$\\
$L_5$ & $\{P,B,C_2,W\}$ & $1$                & $\frac{\epsilon+i\gamma}{1+t^2}$              & $\frac{2h_1+i(2h_2-t)}{2(1+t^2)}$ &  \\ [1ex] % [1ex] adds vertical space
\hline %inserts single line
\end{tabular}
\label{algebras} % is used to refer this table in the text
\end{table}
$\gamma, h_1, h_2$ are constants and $\epsilon=\pm1$. Wave function written in the polar form \eqref{psi},
the basis elements for the symmetry algebras are given by
\begin{equation}
T=\partial_t, \quad P=\partial_x, \quad W=\partial_{\omega}, \quad
B=t\partial_x+\frac{1}{2}\,x\partial_{\omega}\quad
\end{equation}
and
$$C_1=t^2\partial_t+xt\partial_x-\frac{1}{2}t\rho\partial_{\rho}+\frac{1}{4}x^2\partial_{\omega},\quad
D_1=2t\partial_t+x\partial_x-\frac{1}{2}\rho\partial_{\rho}$$ for
$L_1$,
$D_2=t\partial_t+\frac{1}{2}x\partial_x-\frac{1}{2}\rho\partial_{\rho}$
for $L_3$ and $L_4$, and
$C_2=(1+t^2)\partial_t+xt\partial_x+\frac{1}{4}x^2\partial_{\omega}$
for the algebra  $L_5$. We note that $L_1$ is a non-solvable and
$L_2$ is a nilpotent algebra and the other three are solvable and
non-nilpotent. In addition, $L_1$ and $L_3$ are decomposable
whereas the others are not.

These canonical equations do not pass  the Painlevé test for PDEs,
therefore they are not integrable and will be the main subject of
this study. We are going to apply two different methods: Symmetry reduction
and Painlevé truncated expansions. The first is  to make use of
the one-dimensional subalgebras of the four-dimensional algebras
given in Table \ref{algebras} and the second is  to find a valid
truncated series solution to the equation.

The paper is organized as follows. In Section 2   we find the
group-invariant equations for the canonical equations having four
dimensional symmetry algebras. Section 3 is devoted to the
analysis of the reduced systems and completes the study of the
invariant solutions. In Section 4 we apply the method of truncated
Painlevé expansions to the canonical equations to obtain exact
solutions.

\section{One-dimensional subalgebras and reductions to ODEs}

As we are interested in group invariant solutions  we only need one-dimensional subalgebras. This is the case because we  restrict ourselves to subgroups of the symmetry group having generic orbits of codimension 3 in the space  $\{x,t\}\times \{\rho, \omega\}$.

The classification of one-dimensional subalgebras  under the action of the group of inner automorphisms of the four-dimensinal symmetry groups is a standard one. We do not provide the calculations
leading to the conjugacy inequivalent list of subalgebras. The classification method  can be found in for example in \cite{Olver95,Ovsi,Winternitz1992}.

The main result is that every one-dimensional subalgebra of the symmetry algebra is precisely conjugate to one of the subalgebras given in the Table \ref{t1}.

\begin{table}[ht]\label{t1}
\caption{One-dimensional subalgebras of 4-dimensional algebras under  the adjoint action of the full symmetry group} % title of Table
\centering % used for centering table
\begin{tabular}{l l l l  } % sola dayal{\i} s\"{u}tunlar (4 columns)
\hline\hline %inserts double horizontal lines
Algebra  &   &  Subalgebra  & $a,b,c\in\mathbb{R}$, $\epsilon_1=\mp1$   \\ [0.5ex] % inserts table
%heading
\hline % inserts single horizontal line
$L_1$ & $L_{1.1}=\{T+C_1+aW\}$         & $L_{1.2}=\{D_1+bW\}$ & $L_{1.3}=\{T+cW\}$  \\ % inserting body of the table
\hline
$L_2$ & $L_{2.1}=\{P\}$              & $L_{2.2}=\{T+aW\}$ & $L_{2.3}=\{B+bT\}$  \\
      & $L_{2.4}=\{W\}$              &                    &                      \\
\hline
$L_3$ & $L_{3.1}=\{T\}$              & $L_{3.2}=\{P\}$    & $L_{3.3}=\{T+\epsilon_1W\}$ \\
      & $L_{3.4}=\{P+\epsilon_1W\}$  & $L_{3.5}=\{D_2+aW\}$ &$L_{3.6}=\{T+\epsilon_1P+bW\}$\\
      & $L_{3.7}=\{W\}$              &                    &                               \\
\hline
$L_4$ & $L_{4.1}=\{P\}$              & $L_{4.2}=\{B\}$    & $L_{4.3}=\{P+\epsilon_1B\}$    \\
      &$L_{4.4}=\{D_2+aW\}$            &  $L_{4.5}=\{W\}$   &                                \\
\hline
$L_5$ & $L_{5.1}=\{P\}$              & $L_{5.2}=\{B\}$    & $L_{5.3}=\{C_2+aW\}$              \\ [1ex] % [1ex] adds vertical space
\hline %inserts single line
\end{tabular}
\label{table:nonlin} % is used to refer this table in the text
\end{table}

Using these subalgebras, we  perform the reductions leading to the ODEs. We exclude the subalgebras whenever the invertibility requirement is violated. This is only case for the gauge symmetry $W$ and it does not lead to any group-invariant solutions.    We first write the wave function in the form \eqref{psi} and obtain \eqref{VCS} as a system of two real second order nonlinear PDEs, given by
\begin{subequations}\label{reduce}
\begin{eqnarray}
   \label{reduce1}  -\rho \, \omega_t+f_1 (\rho_{xx}-\rho \omega_x^2)-f_2 (2\rho_x \omega_x+\rho\omega_{xx})+g_1\rho^3+h_1\rho&=&0, \\
   \label{reduce2}  \rho_t+f_2 (\rho_{xx}-\rho \omega_x^2)+f_1 (2\rho_x \omega_x+\rho\omega_{xx})+g_2\rho^3+h_2\rho&=&0.
\end{eqnarray}
\end{subequations}

 In this system  coefficient functions with indices $1,2$ are real and imaginary parts of $f,g$ and $h$. They are all functions of $x$ and $t$. For example, if we  would like to see the system  for the algebra $L_1$, looking at the Table \ref{algebras} we simply replace $h_1(x,t)$ of \eqref{reduce1} by $\frac{h_1}{x^2}$ and for the algebra $L_5$ by $\frac{h_1}{1+t^2}$, this time $h_1$ being a constant.

Invariant surface condition for a specific  subalgebra gives the
similarity variable for the functions $\rho$ and $\omega$. Use of
this variable in \eqref{reduce} therefore reduces the number of
independent variables in the system  from two to one, converting
it to a system of ODEs. These nonlinear systems of ODEs arise as
first or second order nonlinear equations.

In the following, each time we encounter a first order equation, we provide its solution right away. The task is harder for the second order equations, because we have to decouple the reduced system of equations before treating them. For specific values of the constants appearing in the reduced equations, we were able to succeed in decoupling the systems and left the search for solutions to the Section 3.

\subsection{Non-Solvable Algebra ${L_1=\{T,D_1,C_1,W\}}$}
Commutators for the basis elements of the four-dimensional algebra $L_1$ satisfy
\begin{equation}\label{}
[T,D_1]=2T, \quad [T,C_1]=D_1, \quad [D_1,C_1]=2C_1
\end{equation}
with $W$ being the center element, that is, commuting with all the other elements. The
algebra has the direct sum structure
\begin{equation}\label{}
L_1=\Sl(2,\mathbb{R})\oplus \{W\}.
\end{equation}
Representative equation of the algebra is
\begin{equation}\label{}
i\psi_t+\psi_{xx}+(\epsilon+i \gamma )\frac{1}{x} |\psi|^2 \psi+(h_1+i h_2)\frac{1}{x^2} \psi=0
\end{equation}
with the real constants $\epsilon=\mp1, \gamma, h_1, h_2$. In
polar variables it takes the form of a system
\begin{subequations}\label{red1}
\begin{eqnarray}
   \label{red1a}  -\rho \, \omega_t+\rho_{xx}-\rho \omega_x^2+\frac{\epsilon}{x}\rho^3+\frac{h_1}{x^2}\rho&=&0, \\
   \label{red1b}  \rho_t+2\rho_x \omega_x+\rho\omega_{xx}+\frac{\gamma}{x}\rho^3+\frac{h_2}{x^2}\rho&=&0.
\end{eqnarray}
\end{subequations}
\subsubsection{Subalgebra ${L_{1.1}=\{T+C_1+aW\}}$}
Invariance under the subalgebra $L_{1.1}$ implies that the
solution has the form
\begin{equation}\label{11}
    \psi(x,t)=\frac{M(\xi)}{\sqrt{x}} \exp\left[\ii\Big(a \arctan t+\frac{x^2
    t}{4(1+t^2)}+P(\xi)\Big)\right],\quad \xi=\frac{x^2}{1+t^2}.
\end{equation}
We substitute $\rho(x,t)=\frac{M(\xi)}{\sqrt{x}}$ and $\omega(x,t)=a \arctan t+\frac{x^2
    t}{4(1+t^2)}+P(\xi)$ in \eqref{red1} and obtain the reduced system of equations satisfied by $M(\xi)$ and $P(\xi)$:
\begin{subequations}\label{11ab}
\begin{eqnarray}
   \label{11ai}  &&\epsilon M^3+\Big(\frac{3-\xi^2}{4}-a \xi+h_1\Big)M-4\xi^2M P'^2+4 \xi^2 M''=0, \\
   \label{11bi}  &&\gamma M^3+h_2M+8\xi^2M'P'+4 \xi^2 M P''=0.
\end{eqnarray}
\end{subequations}
We first need to decouple these equations to solve for the functions
$M$ and $P$. If \eqref{11bi} is multiplied by $M$ and written as
\begin{equation}
     \label{11b}  \gamma M^4+h_2M^2+4\xi^2(M^2 P')'=0,
\end{equation}
it is seen that an integral of \eqref{11b} can be obtained for
three different cases of the constants.

\noindent (i) The case $\gamma=0$, $h_2\neq0$.

We define $Y=Y(\xi)$ in \eqref{11b} as
\begin{equation}
(M^2P')'=-\frac{h_2 M^2}{4 \xi^2}=Y',
\end{equation}
which then easily integrates to
\begin{equation}
M^2P'=Y+C, \quad C=\rm{const.}
\end{equation}
so that
\begin{equation}\label{M2}
M^2=-\frac{4}{h_2}\xi^2 Y'.
\end{equation}
Hence the expression
\begin{equation}
P'=-\frac{h_2 (Y+C)}{4 \xi^2
Y'}
\end{equation}
obtained in terms of $Y$ has to be substituted into \eqref{11ai}. But
the other terms in \eqref{11ai} must also be  expressed in terms
of $Y$, which is seen to be possible if one considers \eqref{M2}.
As a result we obtain a third order nonlinear ordinary
differential equation  from \eqref{11ai}:
\begin{equation}\label{311}
Y'Y'''-\frac{1}{2}Y''^2+\frac{2}{\xi}Y'Y''+\frac{1}{2\xi^2}\Big(\frac{3-\xi^2}{4}-a
\xi+h_1\Big)Y'^2-\frac{2\epsilon}{h_2}Y'^3-\frac{h_2^2}{8\xi^4}(Y+C)^2=0.
\end{equation}
From a solution of this equation, one can find the functions $M,P$
of \eqref{11} as
\begin{equation}
M(\xi)=2\xi\sqrt{-\frac{1}{h_2} Y'}\,, \qquad
P(\xi)=-\frac{h_2}{4}\int \frac{Y+C}{\xi^2 Y'} \, d\xi.
\end{equation}

\noindent (ii) The case $\gamma\neq0$, $h_2=0$.

This time the function $Y$ is defined in  \eqref{11b} as
\begin{equation}
(M^2P')'=-\frac{\gamma M^4}{4 \xi^2}=Y'
\end{equation}
leading to the integral
\begin{equation}\label{P}
M^2P'=Y+C, \quad C=\rm{const.}
\end{equation}
so that we have
\begin{equation}\label{M22}
M^4=-\frac{4}{\gamma}\,\xi^2\, Y'.
\end{equation}
From \eqref{11ai} the following third order ODE for
$Y$ is obtained:
\begin{equation}\label{Y12}
Y'Y'''-\frac{3}{4}Y''^2+\frac{1}{\xi}Y'Y''-\Big(\frac{1-4h_1}{4\xi^2}+\frac{a}{\xi}+
\frac{1}{4}\Big)Y'^2+\frac{\gamma}{\xi^2}(Y+C)^2Y'+\frac{2\epsilon}{\xi}Y'^2\sqrt{-\frac{Y'}{\gamma}}=0.\\
\end{equation}
 For a solution  $Y(\xi)$ of this equation  $M$ and $P$ are
 going to be evaluated from
\begin{equation}
M(\xi)=\Big(-\frac{4}{\gamma}\xi^2Y'\Big)^{1/4}\, , \qquad
P(\xi)=\frac{1}{2}\int\sqrt{-\frac{\gamma}{Y'}} \frac{Y+C}{\xi}\,
d\xi.
\end{equation}

\noindent (iii) The case $\gamma=h_2=0$.

In this case we can easily decouple the reduced system of
equations. Integration of  \eqref{11b} gives
\begin{equation}
M^2P'=C,  \qquad P(\xi)=\int \frac{C}{M^2}\,d\xi, \quad
C=\rm{const.}
\end{equation}
and from  \eqref{11ai} we obtain the equation for $M$
\begin{equation}\label{211}
M''=-\frac{\epsilon}{4\xi^2}M^3+\frac{1}{4\xi^2}\Big(\frac{\xi^2-3}{4}+a\xi-h_1\Big)M+C^2M^{-3}.
\end{equation}

\subsubsection{Subalgebra ${L_{1.2}=\{D_1+bW\}}$}

Group-invariant solutions for subalgebra $L_{1.2}$ will have the form
\begin{equation}\label{12}
    \psi(x,t)=\frac{M(\xi)}{\sqrt{x}} \exp\Big[\ii\big(b\ln x+P(\xi)\big)\Big],\quad
    \xi=\frac{x^2}{t}.
\end{equation}
It is straightforward to see that  $M(\xi)$ and $P(\xi)$ must satisfy
\begin{subequations}
\begin{eqnarray}
   \label{12ai}  \epsilon M^3+\Big(\frac{3}{4}-b^2 +h_1\Big)M+\xi(\xi-4b)M P'-4\xi^2MP'^2+4 \xi^2 M''&=&0, \\
   \label{12bi}  \gamma
   M^3+(h_2-2b)M+\xi(4b-\xi)M'+8\xi^2M'P'+4\xi^2MP''&=&0.
\end{eqnarray}
\end{subequations}
If we multiply \eqref{12ai}
 by $M$, we can write it in the form
\begin{equation}
     \label{12b}  \gamma M^4+h_2M^2+\xi^2\left[M^2\Big(\frac{2b}{\xi}-\frac{1}{2}+4P'\Big)\right]'=0.
\end{equation}
Integration of \eqref{12b} is possible in
three different cases.

\noindent (i) The case $\gamma=0$, $h_2\neq0$.

The function $Y(\xi)$ is defined in \eqref{12b} as

\begin{equation}\label{}
    \left[M^2\Big(\frac{2b}{\xi}-\frac{1}{2}+4P'\Big)\right]'=-h_2\frac{M^2}{\xi^2}=Y'.
\end{equation}
We integrate to find
\begin{equation}\label{P12i}
    P'=\frac{1}{8}-\frac{b}{2\xi}-\frac{h_2}{4\xi^2}\frac{Y+C}{Y'}
\end{equation}
and with the equality
\begin{equation}\label{M12i}
    M^2=-\frac{1}{h_2}\,\xi^2\,Y'
\end{equation}
\eqref{12ai} can be completely expressed in terms of $Y$
\begin{equation}\label{312}
   Y'Y'''-\frac{1}{2}Y''^2+\frac{2}{\xi}Y'Y''+\frac{1}{8}\Big(\frac{3+4h_1}{\xi^2}-\frac{2b}{\xi}+\frac{1}{4}\Big)Y'^2-\frac{\epsilon}{2h_2}Y'^3-\frac{h_2^2}{8\xi^4}(Y+C)^2=0.
\end{equation}
If a solution $Y$ of this equation is known, then $P$ and $M$ can
be obtained from \eqref{P12i} and \eqref{M12i}.

\noindent (ii) The case $\gamma\neq0$, $h_2=0$.

In \eqref{12b} the function  $Y(\xi)$ is defined by
\begin{equation}\label{2.29}
    \left[M^2\Big(\frac{2b}{\xi}-\frac{1}{2}+4P'\Big)\right]'=-\gamma\frac{M^4}{\xi^2}=Y'.
\end{equation}
We integrate \eqref{2.29} equation to obtain
\begin{equation}\label{P12ii}
    P'=\frac{1}{8}-\frac{b}{2\xi}+\frac{Y+C}{4\xi}\sqrt{-\frac{\gamma}{Y'}}
\end{equation}
and hence
\begin{equation}\label{M12ii}
    M^4=-\frac{1}{\gamma}\xi^2 \, Y'
\end{equation}
of which substitution in \eqref{12ai} gives rise to an equation in
terms of $Y$:
\begin{equation}\label{}
Y'Y'''-\frac{3}{4}Y''^2+\frac{1}{\xi}Y'Y''+\Big(\frac{4h_1-1}{4\xi^2}-\frac{b}{2\xi}+
\frac{1}{16}\Big)Y'^2+\frac{\gamma}{4\xi^2}(Y+C)^2Y'+\frac{\epsilon}{\xi}Y'^2\sqrt{-\frac{Y'}{\gamma}}=0.\\
\end{equation}
 $P$ and $M$ are going to be found from
\eqref{P12ii} and \eqref{M12ii}.

\noindent (iii) The case  $\gamma=h_2=0$.

If \eqref{12b} is integrated once, we get
\begin{equation}\label{}
    P'=\frac{C}{M^2}+\frac{1}{8}-\frac{b}{2\xi}
\end{equation}
and substitution of  this result in \eqref{12ai}
leaves us with the second order equation for $M$:
\begin{equation}\label{212}
   M''=C^2 M^{-3}-\frac{\epsilon}{4\xi^2}M^3-\frac{1}{16
   \xi^2}(3+4h_1-2b\xi+\frac{1}{4}\xi^2)M.
\end{equation}

\subsubsection{Subalgebra ${L_{1.3}=\{T+cW\}}$}

Invariance under the subalgebra  $L_{1.3}$ implies that the
solution will have the form
\begin{equation}\label{13}
    \psi(x,t)=M(x) \exp\Big[\ii\big(ct+P(x)\big)\Big]
\end{equation}
and here $M(\xi)$,  $P(\xi)$ satisfy the
system
\begin{subequations}
\begin{eqnarray}
   \label{13ai}  &&\epsilon x M^3+(h_1-c x^2)M-x^2 M P'^2+x^2 M''=0, \\
   \label{13bi}  &&\gamma x M^3+h_2M+2 x^2M'P'+ x^2 M P''=0.
\end{eqnarray}
\end{subequations}
Similarly, \eqref{13bi} can be arranged as
\begin{equation}
     \label{13b}  \gamma x M^4+h_2M^2+x^2\left(M^2P'\right)'=0
\end{equation}
and with  arguments similar to the preceding algebras we obtain
the following results.

\noindent (i) The case $\gamma=0$, $h_2\neq0$.

$M(x)$ and $P(x)$ are found from
\begin{equation}\label{13}
    M(x)=\left(-\frac{1}{h_2} x^2Y'\right)^{1/2}, \qquad P(x)=-h_2
    \int \frac{Y+C}{x^2 Y'}\,dx.
\end{equation}
Here $Y(x)$ satisfies a third order equation
\begin{equation}\label{313}
    Y' Y'''-\frac{1}{2}Y''^2+\frac{2}{x} Y' Y''+2\Big(\frac{h_1}{x^2}-c\Big) Y'^2-\frac{2 \epsilon x}{h_2}Y'^3-
\frac{2h_2^2}{x^4}(Y+C)^2=0.
\end{equation}

\noindent (ii) The case $\gamma\neq0$, $h_2=0$.

$M(x)$ and $P(x)$ are to be evaluated from
\begin{equation}\label{13}
    M(x)=\left(-\frac{1}{\gamma}xY'\right)^{1/4}, \qquad P(x)=\int
(Y+C)\sqrt{-\frac{\gamma}{xY'}}\,dx.
\end{equation}
$Y(x)$ is  a solution to the equation
\begin{equation}\label{}
Y'Y'''-\frac{3}{4}Y''^2+\frac{1}{2x}Y'Y''+\Big(\frac{16h_1-3}{4x^2}-4c\Big)Y'^2+\frac{4\gamma}{x}(Y+C)^2Y'
+\frac{4\epsilon}{x}Y'^2\sqrt{-\frac{x}{\gamma}Y'}=0.\\
\end{equation}

\noindent (iii) The case $\gamma=h_2=0$.

 $M(x)$ is the  solution of the equation
\begin{equation}\label{213}
    M''=C^2M^{-3}+\Big(c-\frac{h_1}{x^2}\Big)M-\frac{\epsilon}{x}M^3
\end{equation}
and  $P(x)$ is going to be evaluated from
\begin{equation}\label{}
    P(x)=\int \frac{C}{M^2}\, dx.
\end{equation}

\subsection{Nilpotent algebra ${L_2=\{T,P,B,W\}}$}
Nonzero commutation relation is   $[P,B]=\frac{1}{2}W$. The algebra contains the three dimensional abelian ideal $\{T,P,W\}$. The action of $B$ on this ideal  can be represented by the nilpotent matrix $N$

 \[\left( {\begin{array}{*{20}{c}}
{[P,B]}\\
{[T,B]}\\
{[W,B]}
\end{array}} \right) = N\left( {\begin{array}{*{20}{c}}
P\\
T\\
W
\end{array}} \right),    \qquad N = \left( {\begin{array}{*{20}{c}}
0&0&{1/2}\\
0&0&0\\
0&0&0
\end{array}} \right).\]
In this case the canonical equation has the form
\begin{equation}\label{}
i\psi_t+\psi_{xx}+(\epsilon+i \gamma )|\psi|^2 \psi+i h_2 \psi=0
\end{equation}
with the real constants $\epsilon=\mp1, h_2\neq0$ and $\gamma$.

\subsubsection{Subalgebra ${L_{2.1}=\{P\}}$}

The group-invariant solution of $L_{2.1}$ has the form
\begin{equation}\label{12}
    \psi(x,t)=M(t) \exp\big(\ii P(t)\big)
\end{equation}
and $M,P$ must satisfy
\begin{subequations}\label{21}
\begin{eqnarray}
   \label{21ai}  &&\epsilon M^2-P'=0, \\
   \label{21bi}  &&\gamma M^3+h_2 M+M'=0.
\end{eqnarray}
\end{subequations}
We immediately integrate these equations and find
\begin{subequations}
\begin{eqnarray}
   \label{21a}  &&M(t)=\left(M_1\exp(2 h_2 \, t)-\frac{\gamma}{h_2}\right)^{-1/2},\\
   \label{21b}  &&P(t)=\left\{\displaystyle
\begin{array}{lrl}
 \frac{\epsilon}{2 \gamma}\ln\left(M_1-\frac{\gamma}{h_2}\exp(-2 h_2 \, t)\right)+P_1, && \gamma\neq0
\\[0.4cm]
 P_1-\frac{\epsilon}{2h_2M_1}\exp(2 h_2 \, t), && \gamma=0
\end{array}
\right.
\end{eqnarray}
\end{subequations}
with arbitrary constants $M_1, P_1.$
\subsubsection{Subalgebra ${L_{2.2}=\{T+aW\}}$ }

A solution invariant under the algebra $L_{2.2}$ must be in the following form
\begin{equation}\label{}
    \psi(x,t)=M(x) \exp\Big[\ii \big(at+P(x)\big)\Big].
\end{equation}
Here $M,P$ have to satisfy
\begin{subequations}
\begin{eqnarray}
   \label{22ai}  &&\epsilon M^3-a M-MP'^2+M''=0,\\
   \label{22bi}  &&\gamma M^3+h_2 M+2 M' P'+M P''=0.
\end{eqnarray}
\end{subequations}
For $\gamma=0$ we multiply \eqref{22bi} by $M$ to obtain
\begin{equation}
   \label{22b}  h_2 M^2+(M^2 P')'=0
\end{equation}
and define the function $Y(x)$ in this equation  as
\begin{equation}\label{}
    (M^2 P')'=-h_2 M^2=Y'.
\end{equation}
Substitution of
\begin{equation}\label{P22}
     P(x)=-h_2 \int \frac{Y+C}{Y'} \, dx
\end{equation}
in \eqref{22ai} leads to decoupling of the reduced system
\begin{equation}\label{322}
     Y'Y'''-\frac{1}{2}Y''^2-\frac{2\epsilon}{h_2} Y'^3-2 a Y'^2-2
     h_2^2(Y+C)^2=0.
\end{equation}
$P$ is determined by \eqref{P22} and  $M$ is given by
\begin{equation}\label{}
     M(x)=(-\frac{1}{h_2}Y')^{1/2}.
\end{equation}

\subsubsection{Subalgebra ${L_{2.3}=\{B+bT\}}$}

\noindent (i) The case $b\neq0$.  An invariant solution of
$L_{2.3}$ is obtained in the form
\begin{equation}\label{}
    \psi(x,t)=M(\xi) \exp\Big[\ii \Big(\frac{1}{2 b}x t-\frac{1}{6
    b^2}t^3+P(\xi)\Big)\Big], \qquad \xi=b x-\frac{t^2}{2}.
\end{equation}
Functions $M$ and $P$ are solutions to the system
\begin{subequations}
\begin{eqnarray}
   \label{23ai}  &&\frac{\epsilon}{b^2} M^3-\frac{\xi}{2b^4} M-MP'^2+M''=0,\\
   \label{23bi}  &&\gamma M^3+h_2 M+2 b^2 M' P'+b^2 M P''=0.
\end{eqnarray}
\end{subequations}
We can arrange \eqref{23bi} as
\begin{equation}
    \label{23b}  \gamma M^4+h_2 M^2+b^2 (M^2 P')'=0
\end{equation}
and for $\gamma=0$ define $Y(\xi)$ such that
\begin{equation}\label{}
    (M^2 P')'=-\frac{h_2}{b^2} M^2=Y',
\end{equation}
from which we get
\begin{equation}\label{P222}
     P(\xi)=-\frac{h_2}{b^2} \int \frac{Y+C}{Y'} \, d\xi.
\end{equation}
Hence we have decoupled \eqref{23ai} in the  form
\begin{equation}\label{323}
     Y'Y'''-\frac{1}{2}Y''^2-\frac{2\epsilon}{h_2} Y'^3-\frac{x}{b^4} Y'^2-
     \frac{2 h_2^2}{b^4}(Y+C)^2=0.
\end{equation}
$P$ is obtained from \eqref{P222} and $M$ is given by the formula
\begin{equation}\label{}
     M(\xi)=(-\frac{b^2}{h_2}Y')^{1/2}.
\end{equation}

\noindent (ii) The case $b=0$.
\begin{equation}\label{}
    \psi(x,t)=M(t) \exp\Big[\ii \Big(\frac{x^2}{4t}+P(t)\Big)\Big]
\end{equation}
is the form of the group-invariant solution and the reduced system of equations is
\begin{subequations}\label{23ii}
\begin{eqnarray}
   \label{23aii}  &&\epsilon M^2-P'=0,\\
   \label{23bii}  &&\gamma M^3+(h_2+\frac{1}{2t}) M+M'=0.
\end{eqnarray}
\end{subequations}
We solve this system by standard methods and get
\begin{equation}\label{}
    M(t)=\left( t \exp(2 h_2 t) (M_1+2\gamma\int \frac{\exp(-2 h_2 t)}{t}
    dt)\right)^{-1/2},
    \end{equation}
and
\begin{equation}\label{P23}
     P(t)=\left\{\displaystyle
\begin{array}{lrl}
 \frac{\epsilon}{2\gamma}\ln\left(2\gamma\int\frac{\exp(-2h_2t)}{t}dt+M_1\right)+P_1, && \gamma\neq0
\\[0.4cm]
 \frac{\epsilon}{M_1}\int\frac{\exp(-2h_2t)}{t}dt+P_1, && \gamma=0.
\end{array}
\right.
\end{equation}

In the following we study the reductions for $L_3, L_4, L_5$,
which are solvable non-nilpotent Lie algebras. $L_3$ contains an
abelian ideal. $L_4$ and $L_5$  include a nilpotent ideal.

\subsection{Solvable algebra ${L_{3}=\{T,P,D_2,W\}}$}
The algebra has the abelian ideal $\{T,P,W\}$. Nonzero commutation relations are
\begin{equation}\label{}
    [D_2,T]=T, \quad [D_2,P]=\frac{1}{2}P.
\end{equation}
The algebra has a decomposable structure
\begin{equation}\label{}
    L_3=\{T,P,D_2\}\oplus\{W\}.
\end{equation}
We note the canonical equation
\begin{equation}\label{}
    i\psi_t+(1+i f_2)\psi_{xx}+(\epsilon+i \gamma )|\psi|^2 \psi=0
\end{equation}
with the constants $\epsilon=\mp1$, $f_2\neq0, \gamma$ and proceed to find the reduced ODEs.

\subsubsection{Subalgebra ${L_{3.1}=\{T\}}$}

A  solution of VCNLS invariant under the algebra $L_{3.1}$ must
have the form
\begin{equation}\label{}
    \psi(x,t)=M(x) \exp\big(\ii P(x)\big).
\end{equation}
$M$ and $P$ are found from the following reduced system
\begin{subequations}
\begin{eqnarray}
   \label{31ai}  &&\epsilon M^3-2 f_2 M' P'-M (P'^2+f_2
   P'')+M''=0,\\
   \label{31bi}  &&\gamma M^3+2 M' P'+f_2 M''+M(-f_2 P'^2+P'')=0.
\end{eqnarray}
\end{subequations}
It is possible to arrange these equations as
\begin{subequations}
\begin{eqnarray}
   \label{31a}  && \frac{\epsilon+\gamma f_2}{1+f_2^2}\, M^3-M P'^2+ M''=0, \\
   \label{31b}  &&\frac{\gamma-\epsilon f_2}{1+f_2^2}\, M^4+ (M^2 P')'=0.
\end{eqnarray}
\end{subequations}
Similar to the preceding calculations, we were  able to achieve decoupling for special cases of the parameters as follows.

\noindent (i) The case $\gamma \neq \epsilon f_2$. The function
$Y(x)$ defined as
\begin{equation}\label{31MP}
(M^2 P')'=\frac{\epsilon f_2-\gamma}{1+f_2^2} M^4=Y'
\end{equation}
gives us the relations
\begin{equation}\label{31MP2}
P'=\frac{Y+C}{M^2}, \quad M=\left(\frac{1+f_2^2}{\epsilon f_2-\gamma}Y'\right)^{1/4}.
\end{equation}
When we use these in \eqref{31a}, we obtain a third-order equation for $Y(x)$
\begin{equation}
Y'Y'''-\frac{3}{4}Y''^2+\frac{4(\epsilon+\gamma f_2)}{\sqrt{1+f_2^2}}\sqrt{\frac{Y'}{\epsilon f_2-\gamma}}Y'^2+\frac{4(\gamma-\epsilon f_2)}{1+f_2^2}(Y+C)^2 Y'=0.
\end{equation}
For a solution of this equation one can evaluate $M$ and $P$ from equations \eqref{31MP2}.

\noindent (ii) The case $\gamma=\epsilon f_2$.

$M(x)$ is a solution to the equation
\begin{equation}\label{231}
M''=\frac{C^2}{M^3}-\epsilon  M^3
\end{equation}
and $P(x)$ is given by
\begin{equation}
P(x)=\int \frac{C}{M^2} dx.
\end{equation}

\subsubsection{Subalgebra ${L_{3.2}=\{P\}}$}
Modulus and phase for the group-invariant solution corresponding to the algebra  $L_{3.2}$, which is in the form
$\psi(x,t)=M(t)\exp\big(\ii P(t)\big)$, are found from the system
\begin{subequations}\label{32}
\begin{eqnarray}
     &&\epsilon M^2-P'=0,\\
     &&\gamma M^3+M'=0
\end{eqnarray}
\end{subequations}
as
\begin{subequations}
\begin{eqnarray}
  &&M(t)=\sqrt{2\gamma t+M_1},\\
  &&P(t)=\epsilon (\gamma t^2+M_1 t)+P_1.
\end{eqnarray}
\end{subequations}

\subsubsection{Subalgebra ${L_{3.3}=\{T+\epsilon_1 W\}}$ }
The solution in this case should have the form
\begin{equation}
\psi(x,t)=M(x) \exp\Big[\ii \big(\epsilon_1 t+P(x)\big)\Big],
\end{equation}
where $M, \, P$ satisfy
\begin{subequations}
\begin{eqnarray}
  &&\epsilon M^3-2 f_2 M' P'- (\epsilon_1+P'^2+f_2 P'')M+M''=0,\\
  &&\gamma M^3+2 M' P'+(-f_2 P'^2+P'') M+f_2 M''=0.
\end{eqnarray}
\end{subequations}
In order to decouple these equations we can arrange them as
\begin{subequations}
\begin{eqnarray}
 \label{33a} &&\frac{\epsilon+\gamma f_2}{1+f_2^2} M^3-M P'^2+M''=0,\\
 \label{33b} &&(\gamma-\epsilon f_2)M^4+\epsilon_1 f_2
 M^2+(1+f_2^2)(M^2 P')'=0.
\end{eqnarray}
\end{subequations}
Since $f_2\neq0$, a first integral of \eqref{33b}
can be obtained if $\gamma=\epsilon f_2$. Let  $Y(x)$ be defined as
\begin{equation}
(M^2 P')'=-\frac{\epsilon _1 f_2}{1+f_2^2}M^2=Y'.
\end{equation}
Then we have
\begin{equation}
M^2=-\frac{1+f_2^2}{\epsilon_1 f_2}Y', \quad P'=-\frac{\epsilon _1
f_2}{1+f_2^2} \frac{Y+C}{Y'}
\end{equation}
and thus \eqref{33a} is transformed to an equation  in terms of $Y(x)$
\begin{equation}\label{333}
Y' Y'''-\frac{1}{2}Y''^2-\frac{2\epsilon (1+f_2^2)}{\epsilon_1
f_2}Y'^3-\frac{2 f_2^2}{1+f_2^2}(Y+C)^2=0.
\end{equation}
\subsubsection{Subalgebra ${L_{3.4}=\{P+\epsilon_1 W\}}$}
In this case we have
\begin{equation}
\psi(x,t)=M(t) \exp\Big[\ii \big(\epsilon_1 x+P(t)\big)\Big],
\end{equation}
where
\begin{subequations}\label{34}
\begin{eqnarray}
 \label{34a} &&1-\epsilon M^2+P'=0,\\
 \label{34b} &&\gamma M^3-f_2 M+M'=0.
\end{eqnarray}
\end{subequations}
We find that
\begin{subequations}
\begin{eqnarray}
 &&M(t)=\left(M_1 \exp(-2 f_2 \, t)+\frac{\gamma}{f_2}\right)^{-1/2},\\
 &&P(t)=\left\{\displaystyle
\begin{array}{lrl}
 \frac{\epsilon}{2\gamma}\ln \left(M_1+\frac{\gamma}{f_2}\exp(2 f_2 \, t)\right)-t+P_1, && \gamma\neq0
\\[0.4cm]
 \frac{\epsilon}{2f_2M_1}\exp(2 f_2 \, t)-t+P_1, && \gamma=0.
\end{array}
\right.
\end{eqnarray}
\end{subequations}
\subsubsection{Subalgebra ${L_{3.5}=\{D+a W\}}$}
We will look for the solution in the form
\begin{equation}
\psi(x,t)=\frac{1}{x}M(\xi) \exp\Big[\ii \Big(2 a \ln x+P(\xi)\Big)\Big], \quad
\xi=\frac{x^2}{t}
\end{equation}
and the reduced system for $M,P$ is
\begin{subequations}
\begin{eqnarray}
 \nonumber &&\epsilon M^3+(2+6af_2-4a^2)M-2(1+4 a f_2) \xi M'+\big(\xi^2+2(f_2-4 a)\xi\big)M
 P'\\
  \label{35ai} &&-8 f_2 \xi^2 M' P'-4\xi^2MP'^2+4\xi^2M''-4f_2\xi^2MP''=0,\\
 \nonumber &&\gamma M^3+(2f_2-6a-4a^2f_2)M+(8a-2f_2-\xi)\xi M'-2(1+4af_2)\xi
 MP'\\
  \label{35bi} &&+8\xi^2M'P'-4f_2\xi^2MP'^2+4f_2\xi^2M''+4\xi^2MP''=0.
\end{eqnarray}
\end{subequations}
It is possible to rewrite these equations as
\begin{subequations}
\begin{eqnarray}
 \nonumber &&(\epsilon+\gamma f_2)M^3+2(1-2a^2)(1+f_2^2)M-\big(2(1+f_2^2)\xi+f_2\xi^2\big)M'\\
  \label{35ai} &&+\big(\xi^2-8a(1+f_2^2)\xi\big)M P'+4(1+f_2^2)\xi^2(M''-MP'^2)=0,\\
 \nonumber &&(\gamma-\epsilon f_2)M^4-6a(1+f_2^2)M^2+\big(4a(1+f_2^2)\xi-\frac{\xi^2}{2}\big)(M^2)'\\
  \label{35bi} &&-\big(2(1+f_2^2)\xi+f_2\xi^2\big)M^2P'+4(1+f_2^2)\xi^2(M^2P')'=0
\end{eqnarray}
\end{subequations}
so that they contain second order derivatives in terms of only $M$ or $P$. But unfortunately
we have not been able to proceed further as in the previous algebras.

\subsubsection{Subalgebra ${L_{3.6}=T+\epsilon_1 P+b W}$}
The modulus $M$ and the phase $P$ of the group-invariant solution
\begin{equation}
\psi(x,t)=M(\xi) \exp\Big[\ii \big(b t+P(\xi)\big)\Big], \quad \xi=x-\epsilon_1 t
\end{equation}
satisfy the system
\begin{subequations}
\begin{eqnarray}
 \nonumber   &&(\epsilon+\gamma f_2)M^3-b M-\epsilon_1 f_2 M'+\epsilon_1MP'-(1+f_2^2)MP'^2\\
 \label{36a} &&+(1+f_2^2)M''=0,\\
  \nonumber            &&(\gamma-\epsilon f_2)M^4+b f_2 M^2-\frac{\epsilon_1}{2}(M^2)'-\epsilon_1 f_2 M^2 P'\\
 \label{36b}  &&+(1+f_2^2)(M^2 P')'=0.
\end{eqnarray}
\end{subequations}
Arranging \eqref{36b}  according to the terms $M^2 P'$ and $M^2$
we write
\begin{equation}
(M^2P')'-\frac{\epsilon_1 f_2}{1+f_2^2} (M^2 P')=\frac{1}{1+f_2^2}\left(\frac{\epsilon_1}{2}(M^2)'-b f_2 M^2+(\epsilon f_2-\gamma)M^4\right).
\end{equation}
We multiply this equality by $\exp\big(\frac{-\epsilon_1 f_2}{1+f_2^2}\;\xi\big)$ and define  $Y(\xi)$ as
\begin{equation}
\left[\exp\big(\frac{-\epsilon_1 f_2}{1+f_2^2}\;\xi\big) M^2P'\right]'=\frac{\exp\big(\frac{-\epsilon_1 f_2}{1+f_2^2}\;\xi\big)}{1+f_2^2}\left(\frac{\epsilon_1}{2}(M^2)'-b f_2 M^2+(\epsilon f_2-\gamma)M^4\right)=Y'.
\end{equation}
First we have
\begin{equation}\label{P36}
M^2 P'=\exp\big(\frac{\epsilon_1 f_2}{1+f_2^2}\;\xi\big) (Y+C).
\end{equation}
On the other hand, we need to solve for $M^2$ from
\begin{equation}\label{Ricky}
(M^2)'-2\epsilon_1bf_2 M^2+2\epsilon_1(\epsilon f_2-\gamma)
(M^2)^2=2\epsilon_1(1+f_2^2)\exp\big(\frac{\epsilon_1 f_2}{1+f_2^2}\;\xi\big)\,Y'.
\end{equation}
This equation is of Riccati type in $M^2$ and a special solution
is needed for its integration. Furthermore, it can be linearized
through the transformation $M^2=\frac{1}{2 \epsilon_1(\epsilon
f_2-\gamma)}\frac{U'}{U}$ at the cost of having its order
increased by one:
\begin{equation}
U''-2\epsilon_1 bf_2 \, U'+4(\gamma-\epsilon
f_2)(1+f_2^2)\exp\Big(\frac{\epsilon_1 f_2}{1+f_2^2}\;\xi\Big)\,Y'
\, U=0.
\end{equation}
Still, this equation does not lead to any immediate solution. Instead, it
will be easier to handle \eqref{Ricky}  by the choice $\gamma=\epsilon f_2$
\begin{equation}\label{Ricky2}
\Big(\exp\big(-2\epsilon_1bf_2\xi\big) \;
M^2\Big)'=2\epsilon_1(1+f_2^2)\exp\Big(\epsilon_1f_2\big(\frac{1}{1+f_2^2}-2b\big)\xi\Big)\;
Y'
\end{equation}
and if $b=\frac{1}{2(1+f_2^2)}$ then we can integrate to find
\begin{equation}\label{Ricky3}
M^2(\xi)=2\epsilon_1(1+f_2^2)\exp\big(\frac{\epsilon_1 f_2}{1+f_2^2}\;\xi\big)\Big(Y(\xi)+C\Big).
\end{equation}
A comparison of this result by \eqref{P36} forces
\begin{equation}\label{Ricky3}
P'=\frac{1}{2\epsilon_1(1+f_2^2)}.
\end{equation}
Therefore we could end up with a decoupled equation for $M$ from
\eqref{36a}
\begin{equation}\label{Ricky3}
M''=\frac{\epsilon_1
f_2}{1+f_2^2}M'+\frac{1}{4(1+f_2^2)^2}M-\epsilon M^3.
\end{equation}

On the other hand,  if \eqref{36b} is arranged with the condition $\gamma=\epsilon
f_2$ as
\begin{equation}\label{U}
\left((1+f_2^2)M^2P'-\frac{\epsilon_1}{2}M^2\right)'=\epsilon_1f_2
M^2 P'-bf_2M^2,
\end{equation}
the choice $b=\frac{1}{2(1+f_2^2)}$ even makes it possible to write this
equation in the simpler form
\begin{equation}
U'=\lambda U
\end{equation}
where $U(\xi)=(1+f_2^2)M^2P'-\frac{\epsilon_1}{2}M^2$,
$\lambda=\frac{\epsilon_1f_2}{1+f_2^2}$. The obvious solution
$U(\xi)=\lambda_0 \exp(\lambda\xi)$ with some constant $\lambda_0$
gives us the formula for $P'$
\begin{equation}
P'=\frac{1}{2\epsilon_1(1+f_2^2)}+\frac{\lambda_0
\exp(\lambda\xi)}{1+f_2^2}M^{-2}.
\end{equation}
Substitution of this result in \eqref{36a} gives the decoupled
equation for $M$ as
\begin{equation}\label{236}
M''=\frac{\epsilon_1
f_2}{1+f_2^2}M'+\frac{1}{4(1+f_2^2)^2}M-\epsilon
M^3+\frac{\lambda_0^2 \exp(2\lambda\xi)}{(1+f_2^2)^2}\,M^{-3}.
\end{equation}
This equation reduces to \eqref{Ricky3} for $\lambda_0=0$.

\subsection{Solvable algebra ${L_4}=\{P,B,D_2,W\}$}
This solvable non-nilpotent algebra  is the extension of the   nilpotent three dimensional Lie algebra $\{W,P,B\}$. We
represent the action of $D_2$ on this ideal  by a  matrix $M$
\[\left( {\begin{array}{*{20}{c}}
{[W,D_2]}\\
{[P,D_2]}\\
{[B,D_2]}
\end{array}} \right) = M\left( {\begin{array}{*{20}{c}}
W\\
P\\
B
\end{array}} \right),    \qquad M = \left( {\begin{array}{*{20}{c}}
0 &   0&0\\
0 &1/2 &0\\
0 &0   &-1/2
\end{array}} \right).\]
We note that the algebra is not decomposable and   the representative equation  from Table 1 is
\begin{equation}\label{}
i\psi_t+\psi_{xx}+(\epsilon+i \gamma )|\psi|^2 \psi+i \frac{h_2}{t} \psi=0,
\end{equation}
where $\epsilon=\mp1, h_2\neq0$ and $\gamma$ are real constants.

\subsubsection{Subalgebra ${L_{4.1}=\{P\}}$}
Group-invariant solution is of the form
\begin{equation}
\psi(x,t)=M(t)\exp\big(\ii P(t)\big)
\end{equation}
and the reduced system  is first order
\begin{subequations}\label{41}
\begin{eqnarray}
\label{41a}&&\epsilon M^2-P'=0,\\
\label{41b}&&M'+\frac{h_2}{t}M+\gamma M^3=0.
\end{eqnarray}
\end{subequations}
Solution of this system is elementary
\begin{subequations}
\begin{eqnarray}
&&M(t)=(2\gamma t\ln t+M_1 t)^{-1/2} \\
&&P(t)=\left\{\displaystyle
\begin{array}{lrl}
\frac{\epsilon}{2\gamma}\ln (2 \gamma \ln t+M_1)+P_1, &&
\gamma\neq0
\\[0.4cm]
 \frac{\epsilon}{M_1}\ln t+P_1, && \gamma=0
\end{array}
\right.
\end{eqnarray}
\end{subequations}
for $h_2=1/2$,  whereas
\begin{subequations}
\begin{eqnarray}
&&M(t)=\Big(\frac{2\gamma}{1-2h_2}\,t+M_1 t^{2h_2}\Big)^{-1/2}, \\
&&P(t)=\left\{\displaystyle
\begin{array}{lrl}
 \frac{\epsilon}{2\gamma}\ln(M_1+\frac{2\gamma}{1-2h_2}\,t^{1-2h_2})+P_1, && \gamma\neq0
\\[0.4cm]
 \frac{\epsilon}{M_1 (1-2h_2)}\,t^{1-2h_2}+P_1, && \gamma=0
\end{array}
\right.
\end{eqnarray}
\end{subequations}
for $h_2\neq1/2$.
\subsubsection{Subalgebra ${L_{4.2}=\{B\}}$}
The solution invariant under $B$ has the form
\begin{equation}
\psi(x,t)=M(t)\exp\Big[\ii \Big(\frac{x^2}{4t}+P(t)\Big)\Big].
\end{equation}
The reduced first order system is
\begin{subequations}\label{42}
\begin{eqnarray}
\label{42a}&&\epsilon M^2-P'=0,\\
\label{42b}&&M'+\frac{1+2h_2}{2t}M+\gamma M^3=0;
\end{eqnarray}
\end{subequations}
which can be integrated to give
\begin{subequations}
\begin{eqnarray}
&&M(t)=(M_1\, t^{1+2h_2}-\frac{\gamma}{h_2}\,t)^{-1/2},\\
&&P(t)=\left\{\displaystyle
\begin{array}{lrl}
\frac{\epsilon}{2\gamma}\ln(M_1-\frac{\gamma}{h_2} t^{-2h_2})+P_1,
&& \gamma\neq0
\\[0.4cm]
 -\frac{\epsilon}{2h_2M_1}t^{-2h_2}+P_1, && \gamma=0.
\end{array}
\right.
\end{eqnarray}
\end{subequations}
\subsubsection{Subalgebra ${L_{4.3}=\{P+\epsilon_1 B\}}$}
The corresponding invariant solution is given by
\begin{equation}
\psi(x,t)=M(t)\exp\Big[\ii \Big(\frac{\epsilon_{1} x^2}{4(1+\epsilon_{1} t)}+P(t)\Big)\Big].
\end{equation}
The reduced system becomes
\begin{subequations}\label{43}
\begin{eqnarray}
\label{43a}&&\epsilon M^2-P'=0,\\
\label{43b}&&M'+\left(\frac{h_2}{t}+\frac{\epsilon_1}{2(1+\epsilon_1 t)}\right)M+\gamma M^3=0.
\end{eqnarray}
\end{subequations}
For the case $h_2=1/2$ the system is integrated easily
\begin{subequations}
\begin{eqnarray}
&&M(t)=\left(t (1+\epsilon_1 t) \Big(2 \gamma \ln (\frac{t}{1+\epsilon_1 t})+M_1\Big)\right)^{-1/2},\\
&&P(t)=\left\{\displaystyle
\begin{array}{lrl}
 \frac{\epsilon}{2\gamma}\ln\left(2\gamma\ln(\frac{t}{1+\epsilon_1 t})+M_1\right)+P_1, && \gamma\neq0
\\[0.4cm]
 \frac{\epsilon}{M_1}\ln(\frac{t}{1+\epsilon_1 t})+P_1, && \gamma=0.
\end{array}
\right.
\end{eqnarray}
\end{subequations}
If $h_2\neq1/2$, then integration is possible in terms of the Gauss' hypergeometric function ${}_2F_1(a,b,c;t)$:
\begin{align}
\nonumber   &M(t)=\left(M_1 t^{2h_2}(1+\epsilon_1
t)+\frac{2\gamma}{1-2h_2}t(1+\epsilon_1t)\; {}_2F_{1}(1-2h_2,1,2-2h_2;-\epsilon_1t)\right)^{-1/2},\\
\label{43bi}  &P(t)=\int \epsilon M^2 dt.
\end{align}
We note that for integer values of
$2h_2$ this solution will simplify to elementary functions.

\subsubsection{Subalgebra ${L_{4.4}=\{D+aW\}}$}
A group-invariant solution invariant under the subalgebra $L_{4.4}$ will be of the form
\begin{equation}
\psi(x,t)=\frac{1}{x}M(\xi)\exp\Big[\ii \big(2a\ln x+P(\xi)\big)\Big], \quad
\xi=\frac{x^2}{t}.
\end{equation}
Functions $M,P$ will be solutions of the system
\begin{subequations}
\begin{eqnarray}
\nonumber          &&\epsilon M^3+2(1-2a^2)M-2\xi M'+(\xi^2-8a\xi)M P'-4\xi^2MP'^2\\
\label{44ai}       &&+4\xi^2M''=0,\\
\nonumber          &&\gamma M^4+(h_2\xi-6a)M^2+\big(4a\xi-\frac{\xi^2}{2}\big)(M^2)'-2\xi M^2P'\\
\label{44bi}      &&+4\xi^2(M^2P')'=0.
\end{eqnarray}
\end{subequations}
For $\gamma=0$ we define in \eqref{44bi} the function $Y(\xi)$ as
\begin{equation}
4\xi^2(M^2P')'-2\xi
M^2P'=\big(\frac{\xi^2}{2}-4a\xi\big)(M^2)'+(6a-h_2\xi)M^2=Y'.
\end{equation}
From these relations we find
\begin{equation}\label{1}
M^2P'=\frac{\xi^{1/2}}{4}\Big(\int\xi^{-5/2}Y'd\xi+C\Big)
\end{equation}
and for $h_2=1/4$
\begin{equation}\label{2}
M^2=\frac{2\xi^{3/2}}{\xi-8a}\Big(\int\xi^{-5/2}Y'd\xi+C\Big).
\end{equation}
Thus if $h_2=1/4$ we can obtain from \eqref{1} and \eqref{2} that
\begin{equation}
P(\xi)=\frac{\xi}{8}-a \ln \xi+P_1.
\end{equation}
This special form of $P(\xi)$ is readily seen to satisfy \eqref{44bi} whereas \eqref{44ai} is decoupled to determine $M$ from
\begin{equation}\label{244}
M''=\frac{1}{2\xi}M'-(\frac{1}{64}-\frac{a}{4\xi}+\frac{1}{2\xi^2})M-\frac{\epsilon}{4\xi^2}M^3.
\end{equation}

\subsection{Solvable algebra ${L_{5}=\{P,B,C_2,W\}}$}
$L_5$ is another canonical extension of the nilpotent algebra
$\{W,P,B\}$ to a solvable non-nilpotent indecomposable
four-dimensional algebra. The element $C_2$ acts on the ideal
$\{W,P,B\}$ by the  matrix $M$ as
\[\left( {\begin{array}{*{20}{c}}
{[W,C_2]}\\
{[P,C_2]}\\
{[B,C_2]}
\end{array}} \right) = M\left( {\begin{array}{*{20}{c}}
W\\
P\\
B
\end{array}} \right),    \qquad M = \left( {\begin{array}{*{20}{c}}
0 &   0&0\\
0 &0 &1\\
0 &-1   &0
\end{array}} \right).\]
Thus the  last canonical equation under investigation will be
\begin{equation}\label{}
i\psi_t+\psi_{xx}+\frac{\epsilon+i \gamma}{1+t^2}\, |\psi|^2 \psi+\frac{2h_1+i(2h_2-t)}{2(1+t^2)} \,\psi=0,
\end{equation}
with the constants $\epsilon=\mp1, h_1, h_2$ and $\gamma$.

\subsubsection{Subalgebra ${L_{5.1}=\{P\}}$}
The group-invariant solution  $\psi(x,t)=M(t)\exp\big(\ii P(t)\big)$ will be obtained by solving $M, P$ from the reduced system
\begin{subequations}\label{51}
\begin{eqnarray}
\label{51a}&&\frac{\epsilon}{1+t^2} M^2+\frac{h_1}{1+t^2}-P'=0,\\
\label{51b}&&M'+\frac{2h_2-t}{2(1+t^2)}M+\frac{\gamma}{1+t^2}
M^3=0.
\end{eqnarray}
\end{subequations}
Through the transformation $M=W^{-1/2}$ in the Bernoulli type
equation \eqref{51b} we have
\begin{equation}
\left(\sqrt{1+t^2}\,\exp(-2h_2\arctan t)\, W\right)'=\frac{2\gamma}{\sqrt{1+t^2}}\,\exp(-2h_2\arctan t).
\end{equation}
$W$ can be expressed by a quadrature and hence
\begin{subequations}
\begin{eqnarray}
&&M(t)=\left(\frac{\exp(2h_2\arctan t)}{\sqrt{1+t^2}}\Big(M_1+2\gamma
\int\frac{\exp(-2h_2\arctan t)}{\sqrt{1+t^2}}\,dt\Big)\right)^{-1/2},\\
&&P(t)=\left\{\displaystyle
\begin{array}{lrl}
 h_1\arctan t+\frac{\epsilon}{2\gamma}\ln \left(M_1+2\gamma
\int\frac{\exp(-2h_2\arctan t)}{\sqrt{1+t^2}}\,dt\right)+P_1, &&
\gamma\neq0
\\[0.4cm]
 h_1\arctan t+\frac{\epsilon}{M_1}\int\frac{\exp(-2h_2\arctan t)}{\sqrt{1+t^2}}\,dt+P_1, && \gamma=0.
\end{array}
\right.
\end{eqnarray}
\end{subequations}
We note that by the transformation $t=\tan z$ in the integrands these results can be expressed in terms of hypergeometric functions.

\subsubsection{Subalgebra ${L_{5.2}=\{B\}}$}
The solution has the form $$\psi(x,t)=M(t)\exp\Big[\ii\big(\frac{x^2}{4t}+P(t)\big)\Big],$$
with functions $M,P$ determined from the system
\begin{subequations}\label{52}
\begin{eqnarray}
\label{52a}&&\frac{\epsilon}{1+t^2} M^2+\frac{h_1}{1+t^2}-P'=0,\\
\label{52b}&&M'+\Big(\frac{2h_2-t}{2(1+t^2)}+\frac{1}{2t}\Big)M+\frac{\gamma}{1+t^2}
M^3=0.
\end{eqnarray}
\end{subequations}
A transformation $M=W^{-1/2}$ in equation \eqref{52b} leads to
\begin{equation}
\left(\frac{\sqrt{1+t^2}}{t}\,\exp(-2h_2\arctan t)\, W\right)'=\frac{2\gamma}{t\sqrt{1+t^2}}\exp(-2h_2\arctan t).
\end{equation}
We find $M,P$ as
\begin{subequations}
\begin{eqnarray}
&&M(t)=\left(\frac{t\,\exp(2h_2\arctan t)}{\sqrt{1+t^2}}\Big(M_1+2\gamma\int\frac{\exp(-2h_2\arctan t)}{t\sqrt{1+t^2}}\,dt\Big)\right)^{-1/2},\\
&&P(t)=\left\{\displaystyle
\begin{array}{lrl}
 h_1\arctan t+\frac{\epsilon}{2\gamma}\ln \left(M_1+2\gamma\int\frac{\exp(-2h_2\arctan t)}{t\sqrt{1+t^2}}\,dt\right)+P_1, && \gamma\neq0
\\[0.4cm]
 h_1\arctan t+\frac{\epsilon}{M_1}\int\frac{\exp(-2h_2\arctan t)}{t\sqrt{1+t^2}}\,dt+P_1, && \gamma=0.
\end{array}
\right.
\end{eqnarray}
\end{subequations}
The transformation $t=\tan z$ is also applicable in the integrands leading to hypergeometric solutions.

\subsubsection{Subalgebra ${L_{5.3}=\{C_2+aW\}}$}
Group-invariant solution must have the form
\begin{equation}
\psi(x,t)=M(\xi) \exp\Big[\ii \Big(a \arctan
t+\frac{x^2t}{4(1+t^2)}+P(\xi)\Big)\Big], \qquad \xi=\frac{x}{\sqrt{1+t^2}}
\end{equation}
Substitution of this solution into the original equation ends up with the system
\begin{subequations}
\begin{eqnarray}
&&\epsilon M^3+\big(h_1-a-\frac{\xi^2}{4}\big)M-MP'^2+M''=0,\\
&&\gamma M^4+h_2M^2+(M^2P')'=0.
\end{eqnarray}
\end{subequations}
In the case $\gamma=0, \, h_2\neq0$ decoupling of these equations is possible
\begin{align}\label{353}
&Y'Y'''-\frac{1}{2}Y''^2-\frac{2\epsilon}{h_2}Y'^3+\big(2h_1-2a-\frac{\xi^2}{2}\big)Y'^2-2h_2^2(Y+C)^2=0,\\
&M=(-\frac{1}{h_2}Y')^{1/2}, \quad P'=-h_2\frac{Y+C}{Y'}.
\end{align}
If $\gamma=h_2=0$ then we have a second order nonlinear ODE for $M$
\begin{subequations}\label{253}
\begin{eqnarray}
&&M''=C^2M^{-3}+(\frac{\xi^2}{4}+a-h_1)M-\epsilon M^3,\\
&&P'=CM^{-2}.
\end{eqnarray}
\end{subequations}

\section{Analysis of the reduced equations}

In Table \ref{numbers} we refer to the numbers of the reduced system of equations of first order besides the second and third order equations obtained through the decoupling task for some special values of the arbitrary parameters. Among the third order equations we only included those which are polynomials in the derivative $Y'$, e.g. \eqref{Y12} is not in the list. We have expressed the solutions of first order equations in the preceding Section. This part of the work will be devoted to the study of solutions of the second and third order equations.

\begin{table}[h]
\caption{Equations under study} % title of Table
\centering % used for centering table
\begin{tabular}{c l} % centered columns (4 columns)
\hline\hline %inserts double horizontal lines
Order & Equation Number  \\ [0.5ex] % inserts table  heading
\hline 1 & \eqref{21}, \eqref{23ii}, \eqref{32}, \eqref{34}, \eqref{41}, \eqref{42}, \\
                  &  \eqref{43}, \eqref{51}, \eqref{52}   \\ % inserting body of the table
\hline 2 &  \eqref{211}, \eqref{212}, \eqref{213}, \eqref{231}, \eqref{236}, \eqref{244}, \eqref{253} \\
\hline 3 & \eqref{311}, \eqref{312}, \eqref{313}, \eqref{322}, \eqref{323}, \eqref{333}, \eqref{353} \\ %inserts single line
\hline
\end{tabular}
\label{numbers} % is used to refer this table in the text
\end{table}

\subsection{Third order equations}

None of the seven third order equations summarized in Table
\ref{numbers} passes the invariant Painlevé test for PDEs.  Since
\eqref{322} and  \eqref{333} do not contain the independent
variable, we can directly lower their order by one if we set
$Y'=W(Y)$. We obtain the following second order equations:

\noindent (i) For equation \eqref{322} with
$\dot{W}=\frac{dW}{dY}$,
\begin{equation}
\ddot{W}=-\frac{1}{2W}\dot{W}^2+\frac{2\epsilon}{h_2}+\frac{2a}{W}+\frac{2h_2^2(Y+C)^2}{W^3}.
\end{equation}

\noindent (ii) For equation \eqref{333},
\begin{equation}
\ddot{W}=-\frac{1}{2W}\dot{W}^2+\frac{2\epsilon(1+f_2^2)}{\epsilon_1f_2}+\frac{2f_2^2}{1+f_2^2} \frac{(Y+C)^2}{W^3}.
\end{equation}
We seeked a first integral to these second order equations in the form
\begin{equation}
A(Y,W)W'^3+B(Y,W)W'^2+F(Y,W)W'+G(Y,W)=I
\end{equation}
for some functions $A, B, F, G$ and a constant $I$, but saw that it is not possible.

For all the third order equations satisfied by $Y=Y(\xi)$ (including \eqref{322} and \eqref{333}) we suggested a first integral of the form
\begin{equation}
A(\xi,Y,Y')\,Y''^2+B(\xi,Y,Y')\,Y''+F(\xi,Y,Y')=I
\end{equation}
with some functions $A, B, F$ and a constant I. A first integral of this type exists only possible for   \eqref{313} for the special values of the constants $c=0$, $h_1=(5+81h_2^2)/36$. The first integral has the form
\begin{align}
&Y''^2+\frac{10}{3x}Y'Y''-\frac{2\epsilon x}{h_2}\,Y'^3+\frac{25+81h_2^2}{9x^2}\,Y'^2+\frac{12h_2^2}{x^3}\,YY'\\
\nonumber &+(\frac{12h_2^2C}{x^3}-\frac{I}{x^{7/3}})\,Y'+\frac{4h_2^2}{x^4}\,(Y+C)^2=0.
\end{align}
We made an unsuccessful attempt to find a further first integral
\begin{equation}
A(x,Y)\,Y'^3+B(x,Y)\,Y'^2+F(x,Y)\,Y'+G(x,Y)=K
\end{equation}
where $K$ is another constant.

\subsection{Second order equations}
Among the second order equations successfully decoupled from the reduced systems,
\eqref{213} passes the P-test  for $h_1=5/36$ and so does \eqref{231} without any condition on the parameters. For the other five equations which do not pass the P-test, we looked for a first integral, assuming $M=M(\xi)$ satisfies an equation of type
\begin{equation}
A(\xi,M)M'^3+B(\xi,M)M'^2+F(\xi,M)M'+G(\xi,M)=I,
\end{equation}
which ended up without any success.

Before proceeding to the solutions of second order equations passing the P-test, we  note  that equation \eqref{236} does not contain the independent variable if we choose $\lambda_0=0$, which means a reduction in order. Indeed, if we set $M'=a_1 W(M)$  with $a_1=\frac{\epsilon_1f_2}{1+f_2^2}$, an Abel equation of the second kind is obtained
\begin{equation}
W(\dot{W}-1)=\frac{a_2}{a_1^2} M-\frac{\epsilon}{a_1} M^3,
\end{equation}
where  $a_2=\frac{1}{4(1+f_2^2)}$.
For $n=\frac{\epsilon_2|f_2|}{\sqrt{1+2f_2^2}}-3$ and $\epsilon_2=\mp1$, $w=w(z)$ a transformation in the parametric form
\begin{equation}
M=z^{\frac{n+2}{2}}w, \quad W=\frac{1}{n+3}\,z^{\frac{n+2}{2}}\,(zw'+\frac{n+2}{2}w)
\end{equation}
converts this equation with $A=-\frac{\epsilon\epsilon_1
f_2(1+f_2^2)}{1+2f_2^2}$ to an equation of  Emden-Fowler type  \cite{Polyanin}
\begin{equation}
w''=Az^nw^3
\end{equation}
which drove the final nail in the coffin.

We close this section with the analysis of equations passing the
P-test. Painlevé and his successors classified second order
differential equations that have at most pole-type singularities
in all their solutions and determined such 50 equivalence classes
together with their representative equations. For details the interested
reader is referred to \cite{Ince}. Since we have two second order
equations passing the P-test, we are going to try to find the
equivalence class to which they may belong.

Transformation of \eqref{231} to the equations numbered PXVIII and PXXXIII in the Painlevé
classification of second order nonlinear ODEs, their first integrals and hence solutions in terms of elementary and elliptic functions in various cases were done in \cite{OG06}.  Since a simple substitution and a careful account of the different cases depending on the constants will suffice to find the results for \eqref{231}, we do not reproduce them here and refer the interested reader to that work.

There remains the treatment of equation \eqref{213}. If we make a
change of the dependent variable as $M=\sqrt{H(x)}, \; H(x)>0$ we have
\begin{equation}
H''=\frac{1}{2H}H'^2+2(c-\frac{h_1}{x^2})H-\frac{2\epsilon}{x}H^2+\frac{2C^2}{H}.
\end{equation}
We apply  a further transformation  $H(x)=\lambda(x) W(\eta(x))$
 and find
\begin{eqnarray}\label{single2}
\begin{split}
    \ddot{W}&=\frac{1}{2W}\dot{W}^2-\frac{1}{\dot{\eta}}\left(\frac{\ddot{\eta}}{\dot{\eta}}
    +\frac{\dot{\lambda}}{\lambda}\right)\dot{W}+
    \frac{1}{\dot{\eta}^2}\left(2(c-\frac{h_1}{x^2})+
    \frac{\dot{\lambda}^2}{2\lambda^2}-\frac{\ddot{\lambda}}{\lambda}\right)W\\
    &-\frac{2\epsilon\lambda
    }{\dot{\eta}^2x}W^2+\frac{2C^2}{\lambda^2\dot{\eta}^2}W^{-1}.
\end{split}
\end{eqnarray}
We will determine $\lambda, \eta$ such that this equation is of
Painlevé-type.

\noindent (i) The case $C\neq0$. If $\lambda, \eta$ and other
constants are chosen as
\begin{equation}
\eta=\eta_0x^{2/3}, \quad \lambda=\lambda_0 x^{1/3}, \quad c<0,
\quad \eta_0=-\Big(\frac{9c}{2}\Big)^{1/3}, \quad h_1=\frac{5}{36}
\end{equation}
an equation quite similar to  PXXXIV is obtained.
\begin{equation}\label{z18}
\ddot{W}=\frac{1}{2W}\dot{W}^2+4\alpha W^2-\eta
W+2\delta^{2}W^{-1},
\end{equation}
where $\alpha=-\Big(\frac{9}{2c^2}\Big)^{1/3}\frac{\epsilon
\lambda_0}{4}$,
$\delta=\frac{3C}{2\lambda_0}\Big(\frac{-2}{9c}\Big)^{1/6}$ and
$\lambda_0$ is arbitrary. By a final transformation
\begin{equation}
2\alpha W=\dot V+V^2+\frac{\eta}{2}
\end{equation}
we see that $V$ satisfies the equation
\begin{equation}\label{P2}
\ddot V=2V^3+\eta V+k, \qquad k=-\frac{1}{2}\pm4\alpha\delta i.
\end{equation}
which is the second  Painlevé transcendent so that
we can express  $V=P_{II}(\eta_0\,x^{2/3})$. Since $W$ is
complex-valued $\lambda_0$ has to be chosen so that the product
$\lambda W$ is real.

\noindent (ii) The case $C=0$. If we choose
\begin{equation}
\eta=\eta_0 x^{2/3}, \quad \lambda=\lambda_0 x^{1/3},
\quad \eta_0=\Big(\frac{9c}{4}\Big)^{1/3}, \quad \lambda_0=-\epsilon\, \Big(\frac{32c^2}{9}\Big)^{1/3},
\quad h_1=\frac{5}{36}
\end{equation}
we arrive at the equation PXX:
\begin{equation}
\ddot{W}=\frac{1}{2W}\dot{W}^2+4W^2+2\eta W.
\end{equation}
Setting $U^2=W$ leads to $P_{II}$ again
$$\ddot U=2U^3+\eta U. $$
We can explicitly give the solution as
\begin{equation}
\psi=\lambda_0^{1/2}x^{1/6}
P_{II}(\eta_0\,x^{2/3})\exp(i (ct+P_0)).
\end{equation}

\section{Solutions by truncation}
In order to investigate Painlevé property for \eqref{VCS} in
\cite{Ozemir10} we wrote the equation with its complex conjugate
as the system
\begin{eqnarray}\label{sys}
\begin{split}
iu_t+f(x,t)u_{xx}+g(x,t) u^2 v+h(x,t) u=0,\\
-iv_t+p(x,t)v_{xx}+q(x,t) u v^2+r(x,t) v=0\\
\end{split}
\end{eqnarray}
and expanded $u, v$ as
\begin{equation}\label{exp}
u(x,t)=\sum_{j=0}^{\infty} \; u_j(x,t)\Phi^{\alpha+j}(x,t), \quad
v(x,t)=\sum_{j=0}^{\infty} \; v_j(x,t)\Phi^{\beta+j}(x,t).
\end{equation}
We obtained the conditions on $f, g, h$ so that all solutions to
VCNLS are in this form. Coefficients for canonical
equations of $4$-dimensional subalgebras do not satisfy the
compatibility conditions of the P-test, therefore they do not have
the Painlevé property. Since the conditions obtained from the
P-test are equivalent to those for having a Lax pair, they are not
integrable \cite{Khawaja10,Brugarino10}. In this case if the
series \eqref{exp} is truncated at an order $j=N$ and plugged in
the equation,  a system of equations for $u_j,
\;\,j\leq N$ and $\Phi$ has to be satisfied. An exact solution is obtained once this system can be solved in a consistent way.

For the Painlevé test to be successful it is required that resonance coefficients $u_j$ corresponding to
the resonance indices $j={-1,0,3,4}$ are arbitrary. This  is true if the compatibility conditions at
resonance levels hold. As we already mentioned, this is not the case for our canonical equations, and we  first
checked whether resonance equations are satisfied at all for some special form of $u_j$'s and $\Phi$. The results were not so promising,
since either no condition for $\Phi$  or
conditions being equivalent to the integrable case can arise. When we were lucky to obtain
a specific form for $\Phi$, conditions other than resonance levels did not hold. Therefore, we could not obtain an exact solution and a
B\"{a}cklund transformation by the truncation approach. However, when we applied the method as it was done in
\cite{YK96}, we were able to obtain nontrivial exact solutions.

As the first step of the Painlevé test the leading orders $\alpha$ and $\beta$
are determined by substitution of $u\sim u_0 \Phi^\alpha$ and $v\sim
v_0\Phi^\beta$ in  \eqref{sys}. Balancing the terms of smallest order requires that
\begin{equation}\label{ab}
\alpha+\beta=-2
\end{equation}
and
\begin{equation}\label{u0v0}
u_0v_0=-\alpha(\alpha-1)\frac{f}{g}\Phi_x^2=
-\beta(\beta-1)\frac{p}{q}\Phi_x^2
\end{equation}
hold. Since the leading orders should be negative integers for the equation to have the Painlevé property, \eqref{ab} implies that
 $\alpha=-1$ and $\beta=-1$. Since we are interested in a case in which the equations do not have the P-property,
we weaken the condition that $\alpha$ and $\beta$ are integers and determine the leading orders by solving  the eqs. \eqref{ab} and \eqref{u0v0} simultaneously.
This will indeed lead to finding exact solutions by truncation approach.

We applied this approach to the canonical equations of  algebras $L_1,\, L_3,\, L_4$ successfully. Overdetermined system of
equations for  $L_2$ and $L_5$ algebras are not compatible and the method fails to apply.

\subsection{Truncation method for the algebra ${L_1}$}

The coefficients for the algebra $L_1$ are  $f=1$, $g=(\epsilon+i\gamma)\frac{1}{x}$, $h=(h_1+ih_2)\frac{1}{x^2}$.
We apply the truncation method to the slightly more general coefficients
\begin{equation}\label{kf}
f=1, \quad  g=(\epsilon+i\gamma)\frac{1}{x^a}, \quad  h=(h_1+ih_2)\frac{1}{x^b}, \quad  a,b\in\mathbb{R}.
\end{equation}
If $a\ne 1$, $b=2$, the equation is  invariant under the 3-dimensional solvable algebra with a basis
$$T=\gen t,\quad D=t\gen t+\frac{x}{2}\gen x+\frac{a-2}{4}\rho\gen \rho,\quad W=\gen \omega.$$ The algebra is extended  for $a=1$.

When we solve \eqref{ab} and \eqref{u0v0} together, for $\gamma \neq 0$  we have
\begin{equation}\label{ab1}
\alpha=-1-i\delta, \quad \beta=-1+i\delta;
\qquad \delta=\frac{-3\epsilon\pm\sqrt{8\gamma^2+9}}{2\gamma}
\end{equation}
and \eqref{u0v0} simplifies as
\begin{equation}\label{u0v01}
u_0v_0=-\frac{3\delta}{\gamma}\,x^a\,\Phi_x^2.
\end{equation}
We truncate the Painlevé expansion at the first order ($j=0$) and suggest that solution has the form
\begin{equation}\label{cozum}
u(x,t)=u_0(x,t)\Phi(x,t)^{-1-i\delta}, \quad
v(x,t)=v_0(x,t)\Phi(x,t)^{-1+i\delta}.
\end{equation}
Putting these expressions in  \eqref{sys}, the terms  $\Phi^{-3\pm i\delta}, \Phi^{-2\pm
i\delta}, \Phi^{-1\pm i\delta}$ will appear. We choose  the coefficients of these terms equal to zero to obtain a system of three equations for $u_0,\; v_0$ and $\Phi$ each of which consists of two equations.
The condition obtained at the level $\Phi^{-3\pm i\delta}$ is   \eqref{u0v01} itself. Coefficients of $\Phi^{-2\pm
i\delta}$ vanish if
\begin{subequations}\label{uvf}
\begin{eqnarray}
\label{uvf1}i\frac{\Phi_t}{2\Phi_x}+\frac{u_{0,x}}{u_0}+\frac{\Phi_{xx}}{2\Phi_x}=0,\\
\label{uvf2}-i\frac{\Phi_t}{2\Phi_x}+\frac{v_{0,x}}{v_0}+\frac{\Phi_{xx}}{2\Phi_x}=0.
\end{eqnarray}
\end{subequations}
Taking the  difference of these two equations and using
\eqref{u0v01} lead to
\begin{equation}\label{Fi}
\Phi(x,t)=\left\{\displaystyle
\begin{array}{lrl}
 \phi_0(t)\, x^{1-\frac{a}{3}}+\phi_1(t), && a\neq3
\\[0.4cm]
\phi_0(t)\,\ln|x|+\phi_1(t), && a=3
\end{array}
\right.
\end{equation}
with arbitrary functions $\phi_0(t)$ and $\phi_1(t)$.
We can also solve $u_0$, $v_0$  by using this form of $\Phi$ in \eqref{uvf}.
\begin{subequations}\label{}
\begin{eqnarray}
\label{uvf1}&&u_0(x,t)=C_1(t) \, x^{a/6} \, \exp\left(-\int\frac{i\Phi_t}{2\Phi_x}dx\right),\\
\label{uvf2}&&v_0(x,t)=C_2(t) \, x^{a/6} \, \exp\left(\int\frac{i\Phi_t}{2\Phi_x}dx\right).
\end{eqnarray}
\end{subequations}
Here $C_1(t)$ and $C_2(t)$ are the arbitrary constants of integration and the expression in the exponential function can be evaluated as
\begin{equation}\label{int}
\int\frac{i\Phi_t}{2\Phi_x}dx=\left\{\displaystyle
                                       \begin{array}{lrl}
                                        \frac{ix^2}{4}\left[\frac{\dot{\phi}_0}{\phi_0}(\ln|x|-\frac{1}{2})+\frac{\dot{\phi}_1}{\phi_0}\right], && a=3
                                        \\[0.4cm]
                                       i(\frac{\dot{\phi}_0}{\phi_0}\frac{x^2}{8}+\frac{\dot{\phi}_1}{\phi_0}\frac{\ln |x|}{4}), && a=-3
                                         \\[0.4cm]
                                       i\left[\frac{\dot{\phi}_0}{\phi_0}\frac{x^2}{4(1-\frac{a}{3})}+\frac{\dot{\phi}_1}{\phi_0}\frac{x^{1+\frac{a}{3}}}{2(1-\frac{a^2}{9})}\right], && a\neq\mp3.
\end{array}
\right.
\end{equation}
Functions $u$ and  $v$ are treated independently in applying the P-test. But in this case, in order that the
system \eqref{sys}  corresponds to \eqref{VCS}, we must have $v^{*}=u$.
This requires that $C_2(t)^{*}=C_1(t)$. As a conclusion of this condition, we get
\begin{equation}\label{C}
|C_1(t)|^2=\left\{\displaystyle
\begin{array}{lrl}
-\frac{3\delta}{\gamma}(1-\frac{a}{3})^2\phi_0^2(t), && a\neq3
\\[0.4cm]
-\frac{3\delta}{\gamma}\phi_0^2(t), && a=3
\end{array}
\right.
\end{equation}
using the special singularity \eqref{Fi} in equation \eqref{u0v01} of order $\Phi^{-3\pm i\delta}$.

Again the constants $\delta$ and $\gamma$ must satisfy
$\frac{\delta}{\gamma}<0$. This means that we have to choose the
negative sign for the formula of $\delta$ in   \eqref{ab1}
\begin{equation}\label{delta}
\delta=\frac{-3\epsilon-\sqrt{8\gamma^2+9}}{2\gamma}.
\end{equation}

We solve the equations obtained by choosing the coefficients of
terms $\Phi^{-1\pm i\delta}$  zero  in various cases depending
on the constants  $a,b$.

\noindent (1.) The case $a=3$.

We solve the equations at order $\Phi^{-1\pm i\delta}$ and find
$b=2$, $h_1=1/4$, $h_2=0$. As a result we have
\begin{equation}\label{}
u_0(x,t)=c\sqrt{x}, \quad \Phi(x,t)=k_0 \ln|x|+k_1,
\end{equation}
where $k_0, k_1$ are real arbitrary constants and  $c\in\mathbb{C}$ with $|c|=k_0\sqrt{-\frac{3\delta}{\gamma}}$.  We write the solution explicitly as
\begin{equation}\label{}
u(x,t)=\frac{c\sqrt{x}}{k_0 \ln|x|+k_1} \, \exp\big(-i\delta\ln(k_0 \ln|x|+k_1)\big).
\end{equation}

\noindent (2.) The case $a=-3$.

For the constants we  have $b=2$, $h_2^2=3+4h_1$. $u_0$ and $\Phi$ are found to be
\begin{equation}\label{}
u_0(x,t)=c\,x^{-\frac{1}{2}+i\frac{h_2}{2}}, \quad \Phi(x,t)=k_0 x^2-2h_2k_0 t+k_1,
\end{equation}
$k_0, \, k_1$ are arbitrary real constants and  $c\in\mathbb{C}$ with the modulus $|c|=2k_0\sqrt{-\frac{3\delta}{\gamma}}$. The solution is given by
\begin{equation}\label{}
u(x,t)=\frac{c}{x^{1/2}\,(k_0 x^2-2h_2k_0 t+k_1)} \, \exp\left(i\ln\frac{x^{h_2/2}}{(k_0 x^2-2h_2k_0 t+k_1)^{\delta}}\right).
\end{equation}

\noindent (3.) The case $a\neq\mp3$.

The system of equations look quite complicated in this case. However, compatibility implies that one of them can be expressed in a relatively simple form
\begin{equation}\label{dd2}
2h_2+\left(\frac{\dot{C_1}}{C_1}+\frac{\dot{C_2}}{C_2}+\frac{a+3}{a-3}\,\frac{\dot{\phi}_0}{\phi_0}\right)x^b+\frac{2a}{a-3}\frac{\dot{\phi}_1}{\phi_0}\, x^{-1+\frac{a}{3}+b}=0.
\end{equation}
In order to complete the analysis, we need to find for different values of $a,b$ the functions  $C_1,\,\phi_0, \, \phi_1$ that will satisfy \eqref{dd2} and the accompanying condition
\begin{align}\label{dd1}
\nonumber&2h_1+\frac{a(a-6)}{18}\,x^{-2+b}+\Big(\frac{\dot{C_1}}{C_1}-\frac{\dot{C_2}}{C_2}\Big)i\,x^b-\frac{9}{2(a-3)^2}\frac{\dot{\phi}_1^2}{\phi_0^2}\,x^{\frac{2a}{3}+b}\\
&+\Big(\frac{3(a-6)}{2(a-3)^2}\frac{\dot{\phi}_0^2}{\phi_0^2}-\frac{3}{2(a-3)}\frac{\ddot{\phi}_0}{\phi_0}\Big)\, x^{2+b}\\
\nonumber&-\frac{18}{(a+3)(a-3)^2}\Big(6\frac{\dot{\phi}_0\dot{\phi}_1}{\phi_0^2}+(a-3)\frac{\ddot{\phi}_1}{\phi_0}\Big)\,x^{-1+\frac{a}{3}+b}=0.
\end{align}

We proceed by considering powers of $x$ in \eqref{dd2}.

\noindent (3.i) The case $0=b=-1+\frac{a}{3}+b$. We require  $a=3$. This case is not possible since we had been able to
find an exact  solution in (1.) for $b=2$.

\noindent  (3.ii) The case $0\neq b=-1+\frac{a}{3}+b$. Since we
must have $a=3$, this case corresponds to (1.).

\noindent  (3.iii) The case $0=b\neq-1+\frac{a}{3}+b$. Equation
\eqref{dd2} requires  $a=0$ or $\dot{\phi}_1=0$.

\noindent (A) The case $a\neq0$. We find $\phi_1(t)=k_1$  as a
constant. Since the coefficient of $x^{-2}$ in \eqref{dd1} must
vanish, we find $a=6$. \eqref{dd1} and  \eqref{dd2} simplifies to
\begin{align}
\label{il}2h_1+i\Big(\frac{\dot{C_1}}{C_1}-\frac{\dot{C_2}}{C_2}\Big)-\frac{\ddot{\phi}_0}{2\phi_0}\, x^{2}=0,\\
\label{ik}2h_2+\frac{\dot{C_1}}{C_1}+\frac{\dot{C_2}}{C_2}+3\frac{\dot{\phi}_0}{\phi_0}=0.
\end{align}
Coefficient of $x^2$ in  \eqref{il} gives
\begin{equation}\label{k2}
\phi_0(t)=k_3 t+k_0
\end{equation}
with constants $k_0,k_3$ and integration of \eqref{ik} results in
\begin{equation}\label{h2}
\exp(2h_2 t)\,C_1\,C_2\,\phi_0^3=s, \quad  s={\rm const}.
\end{equation}
Remembering that $C_2=C_1^{*}$   and considering $|C_1|^2=-\frac{3\delta}{\gamma}\phi_0^2$ from \eqref{C}, we see that  \eqref{k2} and \eqref{h2} are satisfied  only if $h_2=k_3=0$. When we set $C_1(t)=k_0\sqrt{-\frac{3\delta}{\gamma}}\exp\big(iK(t)\big)$, \eqref{il} implies that
$K(t)=h_1t+k_2$.  Eventually we have
\begin{equation}\label{}
u_0(x,t)=k_0\sqrt{-\frac{3\delta}{\gamma}}\, x \,\exp\Big(i(h_1t+k_2)\Big), \quad \Phi(x,t)=k_0 x^{-1}+k_1
\end{equation}
and
\begin{equation}\label{}
u(x,t)=k_0\sqrt{-\frac{3\delta}{\gamma}}\frac{x^2}{k_0+k_1 x} \, \exp\Big(i\big(h_1 t-\delta\ln(k_0 x^{-1}+k_1)+k_2\big)\Big).
\end{equation}

\noindent  (B) The case $a=0$. Through a similar analysis we have
$\phi_0(t)=k_0$, $\phi_1(t)=k_1 t+k_2$,
$C_1(t)=k_0\sqrt{-\frac{3\delta}{\gamma}}\,\exp\big(iK(t)\big)$,
$K(t)=(h_1-\frac{k_1^2}{4k_0^2})t+k_3$ together with the condition
$h_2=0$. Therefore we get
\begin{equation}\label{}
u_0(x,t)=k_0\sqrt{-\frac{3\delta}{\gamma}}\exp\Big[i\Big((h_1-\frac{k_1^2}{4k_0^2})t-\frac{k_1}{2k_0}x+k_3\Big)\Big], \quad \Phi(x,t)=k_0 x+k_1 t+k_2
\end{equation}
and the explicit solution
\begin{equation}\label{}
u(x,t)=\frac{k_0\sqrt{-\frac{3\delta}{\gamma}}}{k_0 x+k_1 t+k_2}\exp\Big[i\Big((h_1-\frac{k_1^2}{4k_0^2})t-\frac{k_1}{2k_0}x-\delta \ln(k_0 x+k_1 t+k_2)+k_3\Big)\Big]
\end{equation}
is obtained.

\noindent  (3.iv) The case $b\neq0=-1+\frac{a}{3}+b$. There is
no solution in the truncated expansion form.

\noindent  (3.v) The case $b\neq0, \, -1+\frac{a}{3}+b\neq0, \,
b\neq-1+\frac{a}{3}+b$. All the coefficients in \eqref{dd2} must
vanish:
\begin{subequations}
\begin{eqnarray}
    &&h_2=0,\\
\label{327b}    &&C_1C_2\phi_0^{\frac{a+3}{a-3}}=s,\\
\label{327c}    &&a\dot{\phi}_1=0.
\end{eqnarray}
\end{subequations}

If we remember that $C_2=C_1^{*}$ and that
$|C_1|^2=-\frac{3\delta}{\gamma}(1-\frac{a}{3})^2\phi_0^2$  from \eqref{C}, \eqref{327b} becomes
\begin{equation}\label{ucc}
-\frac{3\delta}{\gamma}(1-\frac{a}{3})^2\phi_0^{2+\frac{a+3}{a-3}}=s.
\end{equation}
Analysis should be carried out for three different cases according to the constant $a$  under the conditions \eqref{327c} and \eqref{ucc}.

\noindent  (A) The case $a=0 \, (b\neq\{0,1\})$. \eqref{dd2} and
\eqref{dd1} together require $h_1=h_2=0$, $\phi_0(t)=k_0$,
$\phi_1(t)=k_1 t+k_2$,
 $C_1(t)=c\exp(-\frac{\mathrm{ik_1^2}}{4k_0^2}t)$, $|c|^2=-\frac{3\delta}{\gamma}k_0^2$, $c\in\mathbb{C}$. Functions $u_0,\,\Phi$ generate
\begin{equation}\label{}
u_0(x,t)=c\exp\Big(-i\big(\frac{k_1}{2k_0}x+\frac{k_1^2}{4k_0^2}t\big)\Big), \quad \Phi(x,t)=k_0x+k_1t+k_2
\end{equation}
and therefore the exact solution is given by
\begin{equation}\label{}
u(x,t)=\frac{c}{k_0x+k_1t+k_2}\, \exp\Big[-i\Big(\frac{k_1}{2k_0}x+\frac{k_1^2}{4k_0^2}t+
\delta\ln(k_0x+k_1t+k_2)\Big)\Big].
\end{equation}

\noindent  (B) The case $a\neq\{0,1\}$. We find from \eqref{dd2}
and \eqref{dd1} that $\phi_0(t)=k_0$, $\phi_1(t)=k_1$,
$h_1=h_2=0$, $a=6$, $C_1(t)=c$ and
$|c|^2=-\frac{3\delta}{\gamma}k_0^2$. We conclude that
\begin{equation}\label{}
u_0(x,t)=c\, x, \quad \Phi(x,t)=k_0 x^{-1}+k_1
\end{equation}
and
\begin{equation}\label{}
u(x,t)=\frac{cx^2}{k_0 +k_1 x}\,\exp\Big(-i\delta\ln(k_0 x^{-1}+k_1)\Big).
\end{equation}

\noindent (C) The case  $a=1\, (b\neq2/3)$. This last situation
is going to give us the exact solution for the canonical equation
of the algebra $L_1$, namely   the case $a=1$ and $b=2$.

Equation \eqref{dd2} requires  $h_2=0$ and $\phi_1=k_1$.  \eqref{dd1}  is satisfied if and only if $b=2$ and $h_1=5/36$. \eqref{dd1} and \eqref{dd2} take the forms
\begin{align}
\Big(\frac{\dot{C_1}}{C_1}-\frac{\dot{C_2}}{C_2}\Big)\, i x^2+\frac{3}{4}\Big(\frac{\ddot{\phi_0}}{\phi_0}-\frac{5}{2}\frac{\dot{\phi_0}^2}{\phi_0^2}\Big)\,x^4&=0,\\
\Big(\frac{\dot{C_1}}{C_1}-\frac{\dot{C_2}}{C_2}-2\frac{\dot{\phi_0}}{\phi_0}\Big)\,x^2&=0.
\end{align}
We arrive at the functions $\phi_0(t)=(k_0 t+k_2)^{-2/3}$, \, $C_1(t)=c_1 \phi_0(t)$ and  $C_2(t)=c_2 \phi_0(t)$. It is obvious that $c_2=c_1^{*}$ and the condition \eqref{C} requires  $|c_1|^2=\frac{-4\delta}{3\gamma}$, $c_1\in\mathbb{C}$. We sum up the results so far:
\begin{align}\label{}
&u_0(x,t)=\frac{c_1x^{1/6}}{(k_0 t+k_2)^{2/3}}\, \exp\Big(i\frac{k_0x^2}{4(k_0 t+k_2)}\Big),\\
&\Phi(x,t)=(k_0 t+k_2)^{-2/3}\, x^{2/3}+k_1
\end{align}
and hence
\begin{equation}\label{blowed}
u(x,t)=\frac{c_1\, x^{1/6}}{x^{2/3}+k_1(k_0 t+k_2)^{2/3}}\,\exp\Big[i\big(\frac{k_0x^2}{4(k_0 t+k_2)}-\delta\ln(\frac{x^{2/3}}{(k_0 t+k_2)^{2/3}}+k_1)\big)\Big].
\end{equation}

\textbf{Remark.} Since the canonical equation of $L_1$ is
invariant under the action of the group of transformations
$SL(2,\mathbb{R})$, which is the composed action of translation
generated by  $T$, scaling $D_1$ and the conformal transformation
of $C_1$,  the solution
\eqref{blowed} is transformed into a new solution of the canonical equation under this action. Owing to the
invariance property under $C_1$, this transformed solution has a finite
time singularity and therefore was fruitful to study blow-up
profiles. It is exactly this blow-up character  that was used
in \cite{GHO10} to establish the existence of singular behaviours of solutions in the sense of $L_p$ and $L_{\infty}$ norms and  in the distributional sense as well.

\subsection{Truncation method for the algebra ${L_3}$ }

Coefficient functions for the algebra $L_3$ are
\begin{equation}\label{kf3}
f=1+if_2, \quad  g=(\epsilon+i\gamma), \quad  h=0.
\end{equation}
In fact, this constant coefficient case is included in \cite{YK96} but we could not deduce
our results from theirs. Differing from the previous algebra, $f$ contains imaginary part and there
will be a slight modification in the above construction. If we solve
\eqref{ab} and  \eqref{u0v0} together we find the leading orders to be
\begin{equation}\label{ab3}
\alpha=-1-i\delta, \quad \beta=-1+i\delta;
\qquad \delta=\frac{-3(\epsilon+\gamma f_2)\pm\sqrt{9(\epsilon+\gamma f_2)^2+8(\gamma-\epsilon f_2)^2}}{2(\gamma-\epsilon f_2)}
\end{equation}
for $\gamma \neq \epsilon f_2$ and \eqref{u0v0} is equivalent to the condition
\begin{equation}\label{u0v03}
u_0v_0=-\frac{3(1+f_2^2)}{\gamma-\epsilon f_2}\,\delta\,\Phi_x^2.
\end{equation}
We truncate the Painlevé expansion at the first term and propose a solution of the form \eqref{cozum}.
We try to determine $u_0, v_0$ and $\Phi$ by choosing the coefficients of the terms $\Phi^{-3\pm i\delta}, \Phi^{-2\pm
i\delta}, \Phi^{-1\pm i\delta}$ which appear when these ansatze for $u$ and  $v$ are put  in \eqref{sys}.
Equations for order $\Phi^{-3\pm i\delta}$ are equivalent to  \eqref{u0v03}. Terms $\Phi^{-2\pm
i\delta}$ disappear if
\begin{subequations}\label{341}
\begin{eqnarray}
\label{}i\frac{\Phi_t}{\Phi_x}+2(1+if_2)\frac{u_{0,x}}{u_0}+(1+if_2)\frac{\Phi_{xx}}{\Phi_x}=0,\\
\label{}-i\frac{\Phi_t}{\Phi_x}+2(1-if_2)\frac{v_{0,x}}{v_0}+(1-if_2)\frac{\Phi_{xx}}{\Phi_x}=0.
\end{eqnarray}
\end{subequations}
These equations can be arranged as
\begin{subequations}\label{342}
\begin{eqnarray}
\label{uvf31}\frac{i+f_2}{2(1+f_2^2)}\frac{\Phi_t}{\Phi_x}+\frac{u_{0,x}}{u_0}+\frac{\Phi_{xx}}{2\Phi_x}=0,\\
\label{uvf32}\frac{-i+f_2}{2(1+f_2^2)}\frac{\Phi_t}{\Phi_x}+\frac{v_{0,x}}{v_0}+\frac{\Phi_{xx}}{2\Phi_x}=0.
\end{eqnarray}
\end{subequations}
Adding equations \eqref{341} and integrating with respect to $x$ and doing the same for equations
\eqref{342} result in
\begin{subequations}\label{}
\begin{eqnarray}
\label{343a}u_0v_0\Big(\frac{u_0}{v_0}\Big)^{if_2}\Phi_x=F_1(t)\\
\label{343b}\exp\Big(\frac{f_2}{1+f_2^2}\int\frac{\Phi_t}{\Phi_x}dx\Big)u_0v_0\Phi_x=F_2(t)
\end{eqnarray}
\end{subequations}
with some arbitrary functions $F_1$ and $F_2$. Moreover, the fact
that $v_0=u_0^{*}$ brings   \eqref{u0v03} to the form
\begin{equation}
|u_0|^2=-\frac{3(1+f_2^2)}{\gamma-\epsilon f_2}\,\delta\,\Phi_x^2.
\end{equation}
When we put this expression in the absolute form of  \eqref{343a}, we see that $\Phi(x,t)=\phi_0(t)\, x+\phi_1(t)$. This leads to the result that the
exponential term in \eqref{343b} cannot depend on $x$. Therefore we have $\Phi_t=0$, that is,  $\Phi(x,t)=k_0 \, x+k_1$ with the constants $k_0, k_1$. Equations for the order
$\Phi^{-1\pm i\delta}$ become quite simple and we solve them to find $u_0(x,t)=c\in\mathbb{C}$, $|c|^2=-\frac{3(1+f_2^2)}{\gamma-\epsilon f_2}\,\delta\,k_0^2$.
It is necessary to choose the negative sign  for the square root in the formula $\delta$ of  \eqref{ab3}.
The corresponding exact solution  will be
\begin{equation}
u(x,t)=\frac{c}{k_0 \, x+k_1}\, \exp\Big(-i\delta\ln(k_0 \, x+k_1)\Big).
\end{equation}

\subsection{Truncation method for the algebra ${L_4}$}

We repeat the arguments which worked for algebra $L_1$ for the potential $h(x,t)=\frac{h_2}{t}$ of algebra $L_4$ and find
\begin{equation}
u(x,t)=\frac{c}{x+k_0 k_1 t}\, \exp\Big[i\big(\frac{x^2}{4t}-\delta\ln(\frac{x}{k_0 t}+k_1)\big)\Big]
\end{equation}
if $h_2=1/2$.   Here the constant $c\in \mathbb{C}$ must satisfy  $|c|^2=-\frac{3\delta}{\gamma}$, in addition we have   $\delta=\frac{-3\epsilon-\sqrt{8\gamma^2+9}}{2\gamma}$\,.

\end{document}